\documentclass[sigconf]{acmart}

\AtBeginDocument{%
  }

\copyrightyear{2026}
\acmYear{2026}
\setcopyright{cc}
\setcctype{by}
\acmConference[CHI '26]{Proceedings of the 2026 CHI Conference on Human Factors in Computing Systems}{April 13--17, 2026}{Barcelona, Spain}
\acmBooktitle{Proceedings of the 2026 CHI Conference on Human Factors in Computing Systems (CHI '26), April 13--17, 2026, Barcelona, Spain}
\acmDOI{10.1145/3772318.3791889}
\acmISBN{979-8-4007-2278-3/2026/04}

\newcommand{\eat}[1]{}

\usepackage{booktabs}
\usepackage{tabularx}
\usepackage{array}
\usepackage{longtable}
\usepackage{tikz}
\usepackage{graphicx}

\usetikzlibrary{shapes.geometric, arrows, positioning, fit, calc}

\tikzstyle{box} = [rectangle, rounded corners, minimum width=3cm, minimum height=1cm, text centered, draw=black, fill=blue!10, text width=6cm]
\tikzstyle{arrow} = [thick,->,>=stealth]
\tikzstyle{exclusion} = [rectangle, rounded corners, minimum width=2cm, minimum height=0.8cm, text centered, draw=black, fill=red!10, text width=4.5cm]

\begin{document}

\title[Beyond Community Notes: A Framework for Understanding and Building CCS for Social Media]{Beyond Community Notes: A Framework for Understanding and Building Crowdsourced Context Systems for Social Media}

\author{Travis Lloyd}
\email{tgl33@cornell.edu}
\orcid{0009-0009-7393-7105}
\affiliation{%
  \institution{Cornell Tech}
  \city{New York}
  \state{NY}
  \country{USA}
}

\author{Tung Nguyen}
\email{tn375@cornell.edu}
\orcid{0009-0006-6456-4931}
\affiliation{%
  \institution{Cornell Tech}
  \city{New York}
  \state{NY}
  \country{USA}
}

\author{Karen Levy}
\email{karen.levy@cornell.edu}
\orcid{0000-0003-3806-9161}
\affiliation{%
  \institution{Cornell University}
  \city{Ithaca}
  \state{NY}
  \country{USA}
}

\author{Mor Naaman}
\email{mor.naaman@cornell.edu}
\orcid{0000-0002-6436-3877}
\affiliation{%
  \institution{Cornell Tech}
  \city{New York}
  \state{NY}
  \country{USA}
}

\renewcommand{\shortauthors}{Lloyd et al.}

\begin{abstract} 
Social media platforms are increasingly adopting features that display crowdsourced context alongside posts, a technique pioneered by X's Community Notes.
These systems---which we term \textit{Crowdsourced Context Systems} (CCS)---have the potential to reshape the information ecosystem as major platforms embrace them as alternatives to professional fact-checking.
To understand the features and implications of these systems, we conduct a systematic literature review of existing CCS research (n=56) and analyze real-world CCS implementations. 
Based on our analysis, we develop a framework with two components. 
First, we present a theoretical model to conceptualize and define CCS.
Second, we identify a design space encompassing six aspects: participation, inputs, curation, presentation, platform treatment, and transparency. 
We also surface normative implications of different CCS design and implementation choices. 
Our work integrates theoretical, design, and ethical perspectives to establish a foundation for future human-centered research on Crowdsourced Context Systems.
\end{abstract}





\maketitle

\begin{figure*}[]
    \centering
    \fbox{\raisebox{-0.5\height}{\includegraphics[width=0.48\linewidth]{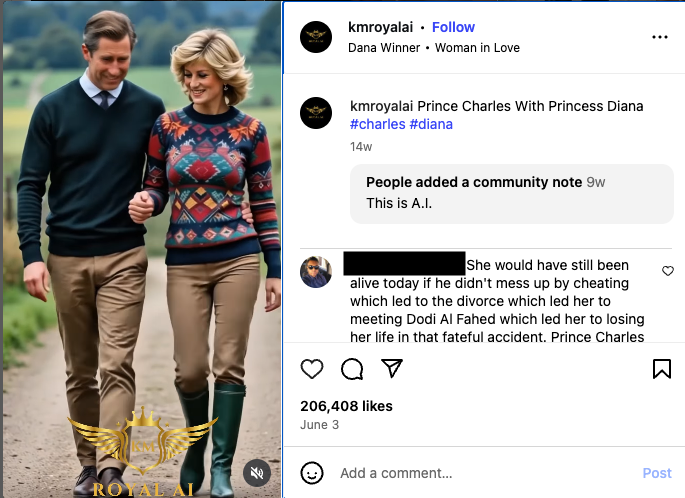}}}\hfill
    \fbox{\raisebox{-0.5\height}{\includegraphics[width=0.48\linewidth]{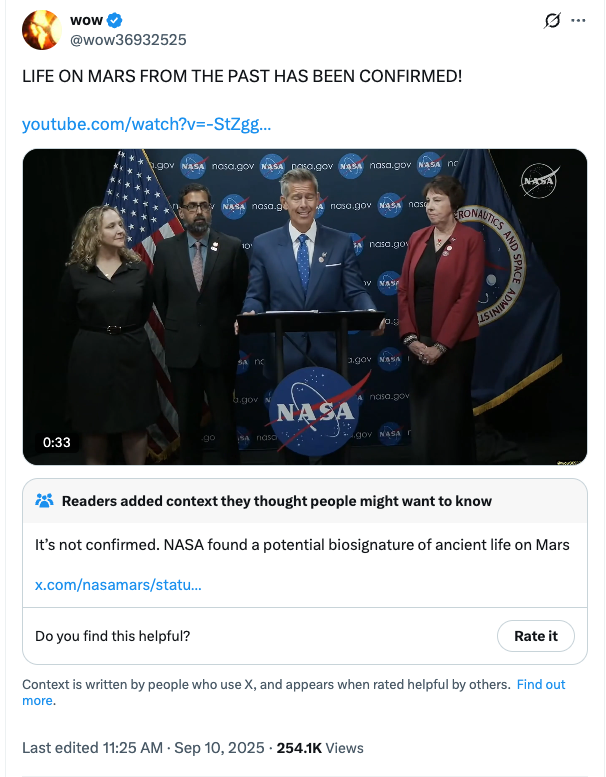}}}
    \caption{Screenshots of crowdsourced notes displayed on Instagram (left) and X (right).}
    \label{fig:screenshots}
    \Description{Two side-by-side screenshots showing Community Notes displayed on different platforms. The left image shows Instagram with a post containing a photo of two people walking, with a Community Note appearing below providing the additional context that the image is AI. The right image shows X (formerly Twitter) with a post about Mars and a Community Note displayed underneath adding the additional context that the post is overstating the claims.}
\end{figure*}

\section{Introduction}
Social media platforms are important sites of entertainment~\cite{herderDigitalJunkfoodSocial2024}, news consumption~\cite{chouakiWhatNewsPeople2024a}, and interpersonal connection~\cite{preeceEmpathicCommunitiesBalancing1999}.
However, these platforms must contend with harmful user-uploaded content, including abuse, spam, and misinformation~\cite{scheuermanFrameworkSeverityHarmful2021}. 
To address misinformation, platforms have tried various content moderation approaches~\cite{grimmelmann2015virtues,lampeSlashdotBurnDistributed2004,jhaverHumanMachineCollaborationContent2019}, from using professional fact-checkers to label or remove false content~\cite{gillespieCustodiansInternetPlatforms2018} to empowering user communities to moderate themselves~\cite{seeringReconsideringSelfModerationRole2020}.
While professional fact-checking has been somewhat successful, it has also evoked backlash and accusations of bias from users and lawmakers, particularly on the political right~\cite{jiangReasoningPoliticalBias2020,johnsonFactsAllegationsPolitical2023}.
Recently, a new approach has emerged: beginning with X/Twitter~\cite{IntroducingBirdwatchCommunitybaseda}, major platforms are increasingly introducing systems that crowdsource ``helpful context'' from users to display alongside potentially problematic posts (see examples in Figure \ref{fig:screenshots}).
These systems, which we term \textit{Crowdsourced Context Systems} (CCS), have seen growing adoption, with some of the world's largest platforms---including Meta (Facebook, Instagram, and Threads), TikTok, and YouTube---all recently launching implementations.
Meta demonstrated the importance of these systems when they announced in January 2025 that their CCS would replace their U.S. fact-checking program~\cite{heatheraMoreSpeechFewer2025}. 

As CCS adoption grows, these systems have the potential to reshape the information ecosystem.
Early studies have produced promising results, finding that CCS can increase user informedness and decrease misinformation spread~\cite{wojcikBirdwatchCrowdWisdom2022b,rollout21,diffusion21}, 
though it remains unclear whether CCS can effectively replace professional fact-checking, as Meta is attempting~\cite{heatheraMoreSpeechFewer2025}.
As CCS develop, novel implementation questions come to the fore, including how to best design these systems to achieve their stated or implicit goals---for example, to provide reliable context quickly enough to prevent the spread of viral misleading content~\cite{rollout21,chuaiCommunitybasedFactcheckingReduces2024a,renaultCollaborativelyAddingContext2024a}.
At this critical juncture in our information ecosystem's development, it is important to understand how CCS function, how to evaluate them, and how to improve their design. 

This paper makes two contributions toward better understanding CCS. 
Our first contribution is a \textbf{systematic literature review} of CCS research.
Our review finds significant work on CCS contributor behavior and the impact of displaying context notes alongside problematic posts.
However, the review also reveals that certain aspects of these systems remain underexplored.
Our second contribution addresses this gap: \textbf{an analysis and comparison of real-world CCS implementations} across major social media platforms.
From this analysis, we create a framework to guide future HCI research on CCS.
Our framework has two parts.
First, we offer a theoretical model that identifies CCS core characteristics and what distinguishes them from prior crowdsourced systems.
Our model posits that while CCS use familiar moderation tools like crowdsourced information and context labels, they are unique in being natively integrated into platforms with the explicit goal of facilitating content interpretation. 
Second, we outline a \textit{design space}~\cite{macleanQuestionsOptionsCriteria1996} for CCS to taxonomize important implementation decisions.
Design spaces are a conceptual HCI tool that guide work on complex socioetechnical systems by structuring design possibilities~\cite{zhangFormFromDesignSpace2024a,leeDesignSpaceIntelligent2024,cardDesignSpaceInput1990}.
We group these decisions into six high-level \textit{aspects} that combine and interact to shape CCS impact: participation, inputs, curation, presentation, platform treatment, and transparency.

Based on our framework, the Discussion section highlights key normative implications of CCS in order to make sense of the ethical stakes involved in their design. 
As these systems aim to shape the information ecosystem, we emphasize three interconnected normative concerns: \textit{user informedness}, \textit{distribution of power}, and \textit{fairness}.
Additionally, we discuss how CCS fit within the broader conversation about content moderation and establish a foundation for
future human-centered research on Crowdsourced Context Systems.

\section{Background}

In January 2021, Twitter (now X) launched a U.S. pilot for the first large-scale CCS, which they called \textit{Birdwatch}, announcing it as ``a community-based approach to misinformation''~\cite{IntroducingBirdwatchCommunitybaseda}.
This system allowed a subset of users to suggest notes to ``provide informative context'' about Tweets (posts) on the platform.
The same users then rated these suggested notes, and the ratings were used as input for a note curation algorithm that aimed to select notes that would be helpful to a broad audience.
In this pilot, these notes were only visible on a separate site, but Twitter announced a plan: ``to make notes visible directly on Tweets for the global Twitter audience, when there is consensus from a broad and diverse set of contributors.''

In March 2022, Twitter expanded the pilot, announcing that: ``a small (and randomized) group of people on Twitter in the US will see Birdwatch notes directly on some Tweets''~\cite{BuildingBetterBirdwatch}.
Twitter was purchased by Elon Musk in October 2022, renamed~X, and soon Birdwatch was renamed \textit{Community Notes}.
In December 2022, the pilot phase ended and users all over the world began seeing selected notes directly alongside Tweets~\cite{BeginningTodayCommunity2022}.
Since then, the system has continued to evolve.
A key feature of X's Community Notes is its transparency: the note curation code is open source\footnote{\url{https://github.com/twitter/communitynotes/tree/main}}, as are the crowdsourced notes and ratings data collected by the system\footnote{\url{https://x.com/i/communitynotes/download-data}}.

In June 2024, YouTube announced a pilot for a similar system, which they called \textit{Notes}~\cite{TestingNewWays}.
In January 2025, Meta announced a pilot for their own implementation, also called \textit{Community Notes}, which would launch on Facebook, Instagram, and Threads~\cite{heatheraMoreSpeechFewer2025}. 
Finally, TikTok announced their own pilot in April 2025, which they called \textit{Footnotes}~\cite{TestingNewFeature}.
At the time of writing, all of these systems have launched, at least as limited pilots.
According to the platforms, all of these systems are based on X's open source note curation algorithm. 
Unlike X's implementation, though, none of the new systems make their code or data public.
In fact, while they serve similar purposes, these systems differ from each other in crucial ways due to platform design decisions. As CCS proliferate on major platforms, the HCI community must define, understand, and reason about them to evaluate their impact and inform better implementations.

\section{Related Work}

While CCS are a relatively new innovation, these systems relate to and touch on multiple bodies of work from the HCI literature, including: 
research on content moderation in public discourse systems, research on fact-checking infrastructures, and research on crowdsourced responses to misinformation.
We describe these areas generally here; later in Section~\ref{lit_review} we provide a detailed review of research that examines specific implementations of CCS. 

In a democracy that values free speech, public participation in civil discussion is considered a social good~\cite{smallPolisEscalarDeliberacion2021}.
One classic concern of HCI research has thus been how to design systems that can support public discussion among users with diverse, possibly conflicting, perspectives~\cite{kripleanSupportingReflectivePublic2012,wohnHowHandleOnline2017}.
CCS are an example of such a system.
Like past public discussion systems, CCS provide a form of \textit{content moderation} that encourages certain types of participation and discourages others~\cite{grimmelmann2015virtues}.
Content moderation can take different approaches: it can be ``top-down,'' where platforms make moderation decisions~\cite{gillespieCustodiansInternetPlatforms2018,gillespieNotRecommendReduction2022}, or ``bottom-up,'' where users do~\cite{seeringReconsideringSelfModerationRole2020,lloyd25,lloydAIRulesCharacterizing2025,fieslerRedditRulesCharacterizing2018}.
Additionally, moderation can be human-powered, or use varying degrees of automation to keep up with large volumes of content~\cite{vaccaroContestabilityContentModeration2021,gorwa20}.
Community Notes, X's CCS implementation, incorporates all of these approaches: it uses a centralized algorithm to aggregate moderation feedback that is sourced from users and, as of late 2025, user-created AI note writers\footnote{\url{https://x.com/CommunityNotes/status/1940132205486915917}}.
X's Community Notes could be seen as an example of a \textit{bridging system}, or an automated system designed to ``increase mutual understanding and trust across divides, creating space for productive conflict, deliberation, or cooperation''~\cite{ovadyaBridgingSystemsOpen2023}.
Indeed, according to its creators, Community Notes was designed to identify crowdsourced content (notes) ``that both inform understanding (decrease propensity to agree with a potentially misleading claim) and are seen as helpful by a diverse population of users''~\cite{wojcikBirdwatchCrowdWisdom2022b}.
CCS thus reflect a long-standing HCI concern with designing systems that support civil discussion through content moderation, adapting these ideas to meet the scale and speed of major social media platforms.

One moderation approach frequently studied in HCI research is the use of professional fact-checking to deal with misinformation~\cite{walterFactCheckingMetaAnalysisWhat2020,youngFactCheckingEffectivenessFunction2018}.
Emerging from journalistic practice, professional fact-checking empowers certain individuals or organizations to verify public claims and publish determinations about their accuracy.
While these techniques can have positive effects, they are time-consuming and require a large amount of professional labor~\cite{hassanDetectingCheckworthyFactual2015,junejaHumanTechnologicalInfrastructures2022}, and thus may struggle to keep up with the speed at which content is published on the internet~\cite{micallefTrueFalseStudying2022}.
As a result, researchers have explored automated ways to scale fact-checking~\cite{warrenShowMeWork2025,guoSurveyAutomatedFactChecking2022,hassanQuestAutomateFactChecking}.
Beyond research into fact-check production, studies have also investigated the effects that fact-checks have on users who encounter them~\cite{eckerPsychologicalDriversMisinformation2022}.
This work has explored both the most effective ways to present fact-checks in order to shift user belief~\cite{claytonRealSolutionsFake2020}, as well as the psychological characteristics that make some users reluctant to engage with them~\cite{tanakaWhoDoesNot2023}.
In addition to issues of efficacy, and the large cost of running such fact-check programs, social media platforms have struggled with allegations and perceptions that their fact-checks are biased against certain political ideologies~\cite{cotterFactCheckingCrisisCOVID192022,jiangReasoningPoliticalBias2020}. 

CCS can be seen as an evolution in content moderation that aims to address the limitations of professional fact-checking by enlisting platform users.
In doing so, these new systems build on a long history of HCI research exploring ways of harnessing the distributed wisdom of a crowd to produce accurate information in networked settings~\cite{pennycookFightingMisinformationSocial2019,heSurveyRoleCrowds2025}.
This research has shown that crowds can effectively perform various tasks, such as sourcing on-the-ground information~\cite{daileyJournalistsCrowdsourcerersResponding2014,agapieCrowdsourcingFieldCase2015}, or fact-checking political claims~\cite{kripleanIntegratingOndemandFactchecking2014}.
Studies have also found that despite being cheaper and faster than professional fact-checking, crowd-based fact-checking can be similarly accurate in certain circumstances ~\cite{allenScalingFactcheckingUsing2021,martelCrowdsCanEffectively2024a,zhaoInsightsComparativeStudy2023}.
However, crowd-based fact-checking is not a silver bullet. 
Users may respond negatively to ``social correction'' from peers~\cite{mosleh21}, the composition of the crowd affects its ability to identify false claims~\cite{godelModeratingMobEvaluating2021}, and crowds representative of the general population can exhibit biases that limit their accuracy~\cite{drawsEffectsCrowdWorker2022}.
Researchers have studied techniques to mitigate these fact-checking biases~\cite{spiekerDiversePerspectivesCan2023} and have explored other ways that crowds can moderate misleading information, such as via digital juries~\cite{fanDigitalJuriesCivicsOriented2020,zhaoEffectsNonEffectsSocial2025}, or via distributed moderation with up/down votes~\cite{lampeSlashdotBurnDistributed2004,epsteinWillCrowdGame2020}.
CCS share much in common with previously explored crowdsourced fact-checking systems, but they have a novel set of processes and structures that lead us to define them as a new class of system. 
Informed by these past bodies of research, we next define what makes these new systems unique.

\section{Generalized Model of Crowdsourced Context Systems for Social Media}

\begin{figure}[]
    \centering
    \includegraphics[width=1\linewidth]{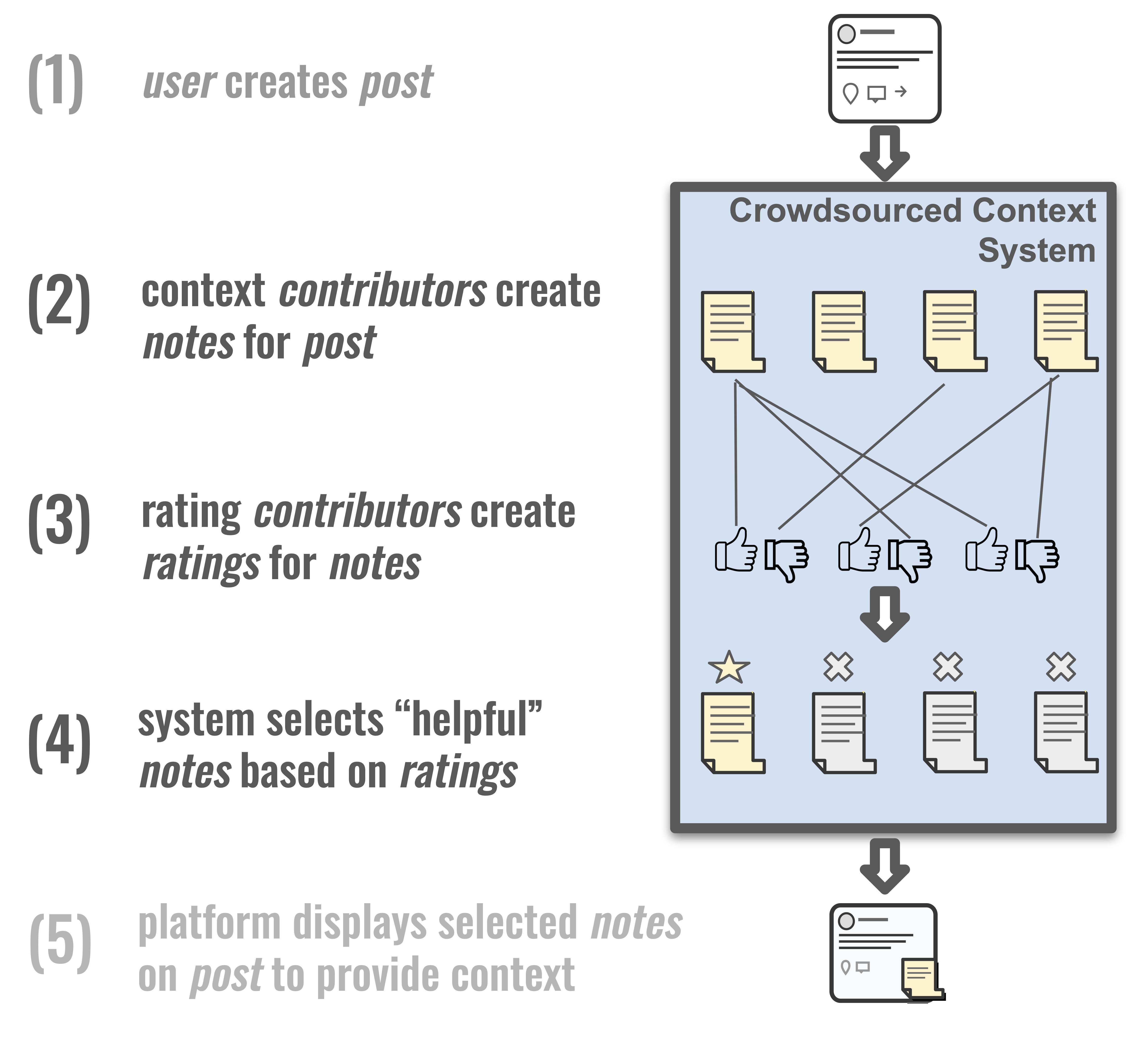}
    \caption{Model of a Crowdsourced Context System. \textit{Users} create \textit{posts}, context \textit{contributors} create \textit{notes} about those \textit{posts}, and ratings \textit{contributors} create \textit{ratings} for those \textit{notes}. Based on these \textit{ratings}, the system selects ``helpful'' \textit{notes}, which are displayed alongside the \textit{post} for \textit{users} to see.}
    \label{fig:model}
    \Description{A flowchart diagram illustrating the Crowdsourced Context System model with 5 numbered steps: (1) User creates post, (2) Context Contributors create Notes for post, (3) Rating Contributors create Ratings for Notes, (4) System selects "Helpful" Notes based on Ratings, and (5) Platform displays selected Notes on post to provide context. The diagram shows the flow between users, the crowdsourced context system, and the final display.}
\end{figure}

We coin the term \textit{Crowdsourced Context System} to describe these social media systems, a novel and important addition to the content moderation toolbox.
Such systems have several core characteristics.
First, CCS rely on contributions \textit{crowdsourced} from users of the platform.
Second, these systems are designed to specifically elicit and identify \textit{context} from the crowd that can help users interpret the original post. 
Crowdsourced context can facilitate interpretation by providing additional information that helps readers assess and interpret the veracity, potential motivations, or broader social or political context of the original post.
This explicit focus on facilitating interpretation differentiates CCS from earlier crowdsourced systems which may have more general aims, such as traditional social media comment sections that emphasize the most engaging responses. 
Third, these systems do not remove user content, but supplement it by displaying additional text, or counterspeech~\cite{pengRescuingCounterspeechBridgingBased2024a}, alongside the original posts.
Finally, these systems are natively integrated into social media platforms. 
Unlike crowd behavior that emerges organically among users, these systems are built and operated by the platforms themselves, giving platform designers the power to influence how the systems function.

Figure~\ref{fig:model} shows a generalized model of a CCS. 
Such systems are composed of several core entities.
\textit{Users} create \textit{posts} on the platform.
\textit{Contributors} then provide context, in the form of text or media snippets called \textit{notes}, which are associated with specific \textit{posts}.
At first these \textit{notes} are only visible to other \textit{contributors}, who assess them and produce \textit{ratings} of their helpfulness.
The platform uses these \textit{ratings} to identify ``helpful'' \textit{notes}, which are then published and displayed alongside the associated \textit{post} everywhere that it appears on the platform, as seen in Figure~\ref{fig:screenshots}.
In implementing this general model, platforms make many individual design decisions, which we highlight and outline below.

\section{Methods}

To better understand CCS and their design possibilities, we performed two distinct reviews.
First, we performed a systematic literature review of existing research on the first large-scale CCS implementation: Twitter's Community Notes. 
Second, we performed inductive thematic analysis and affinity diagramming on a set of CCS design features extracted from documents describing the real-world CCS implementations of several major social media platforms.

\subsection{Systematic Literature Review}

\begin{figure*}
    \centering
    \includegraphics[width=1\linewidth]{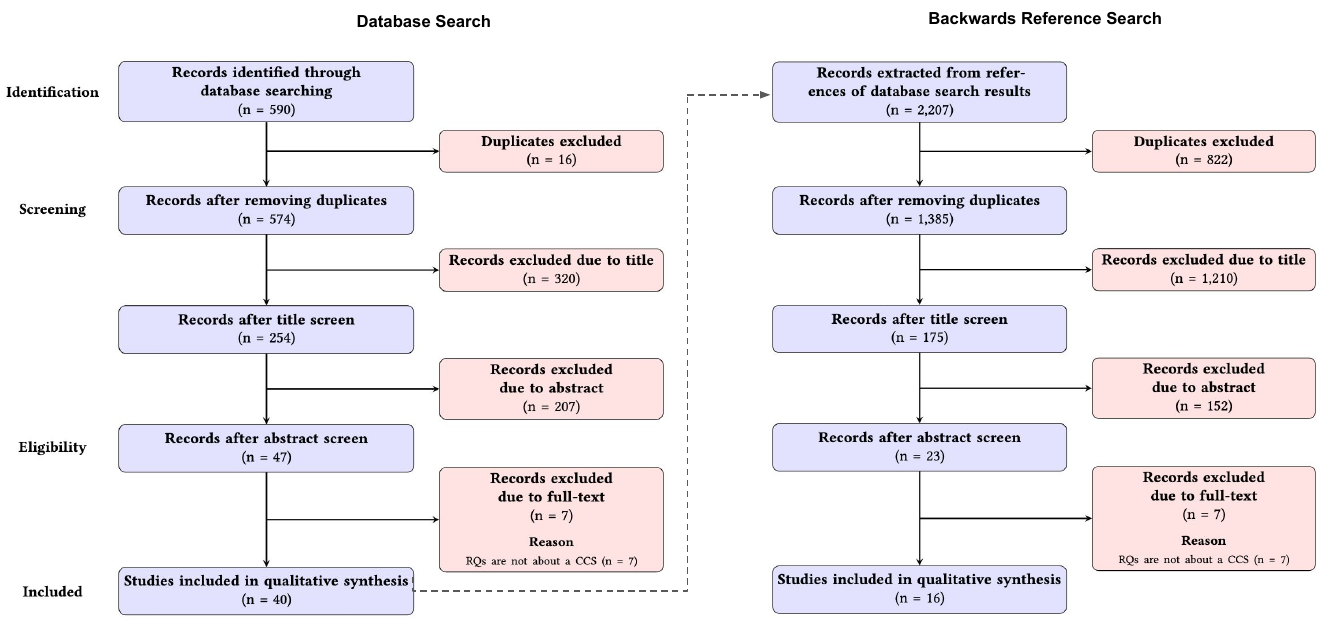}
\caption{PRISMA flow diagram~\cite{pagePRISMA2020Statement2021} for our systematic literature review, showing both the database search and the backwards reference search. We analyzed a final set of papers (n=56) that combined the results of these two searches.}
\label{fig:prisma}
\Description{A PRISMA flow diagram showing the systematic literature review process. It shows two flows side by side: the left is the forwards database search and the right is the backwards reference search. For each flow, the image shows the filtering stages, through identification, screening, and eligibility phases, ultimately resulting in 40 studies from the forward search and 16 from the backwards search included in the qualitative synthesis. Each stage shows the number of records and reasons for exclusions.}
\end{figure*}

In order to synthesize the research community's understanding of CCS, we performed a PRISMA-style systematic literature review~\cite{pagePRISMA2020Statement2021}.
We sourced relevant papers using both a database keyword search and a backwards reference search~\cite{jalaliSystematicLiteratureStudies2012}.
Figure~\ref{fig:prisma} shows a flow diagram summarizing our search and screening process.
Our goal was to identify original research that directly studies CCS implementations.
Since this category of system has not been consistently named prior to our work, we focused our review on studies of the first large-scale implementation of a CCS by a major social media platform: Twitter's Community Notes/Birdwatch.
We thus used ``Birdwatch'' and ``Community Notes'' as keywords and restricted our search to articles that have been written since 2021, when Twitter first launched their implementation.
Note that since this is an emerging technology and the set of articles is limited, we decided to include sources of knowledge that are valued in the HCI community but are sometimes excluded from literature reviews, such as pre-prints and non-full-length conference submissions.
We thus settled on the following inclusion and exclusion criteria:

\textbf{Inclusion criteria:} the study is about a Twitter-style CCS, whether ``in the wild'' or in a controlled setting.

\textbf{Exclusion criteria:} the article is not in English, does not present original research, is only an abstract/a summary of other research, or is not about the functioning of a Twitter-style CCS.

\subsubsection*{Search Strategy} 
We conducted our database search in November 2025. 
We searched seven databases that are  well-known in the fields of HCI and social media research.
We began with six that were used in another recent HCI literature review of social media research~\cite{zhangWhatWeMean2023}: ACM Digital Library, IEEE Xplore, SAGE Journals, ScienceDirect (Elsevier), Taylor \& Francis Online (Tandfonline), and Emerald Insight.
We additionally searched arXiv in order to include preprints, which are common in these fields. 
We searched the full-text, title, abstract, and metadata of articles using the following case-insensitive query string, and limited our results to articles published in 2021 or later: 
\begin{quote}
\centering
\texttt{"community notes" OR "birdwatch"}
\end{quote}
This search produced n=590 results.

\subsubsection*{Title Screening} 
We first screened the search results by title. 
One researcher read the title for each result and removed articles that were either duplicates or were obviously off-topic, such as those that were not about social media systems.  
This step resulted in n=254 remaining articles.

\subsubsection*{Abstract Screening}
We next screened the remaining articles by abstract.
One researcher read the article abstracts and removed redundant articles that only summarized research contained elsewhere in our sample, such as books and book chapters, editorials, abstracts, keynotes, and complete conference proceedings.
We also removed articles that were topically irrelevant, such as those that did not focus on a Twitter-style CCS, or that used data from such a system, but only to answer research questions unrelated to the system.
This step resulted in n=47 remaining articles.

\subsubsection*{Full-Text Screening}
Finally, two researchers read the full text of the remaining articles and screened them further using the inclusion and exclusion criteria.
We met during the process to compare results and discuss any differences or challenging cases until we were in agreement.
This step produced a set of n=40 articles. 

\subsubsection*{Backwards Reference Search}
We then used these 40 articles as the starting set for a backwards reference search and extracted all contained references (n=2,207).
We de-duplicated these references by paper title, resulting in n=1,385 unique papers.
We then performed the same screening steps used in the forward search:
we screened titles, which resulted in n=175 papers, then abstracts, which resulted in n=23 papers, then full texts.
This backwards search identified n=16 relevant papers not found in the forward search.
We combined the sets from both searches into a final set of n=56 papers.

\subsubsection*{Data Extraction}
Two researchers then extracted data from each paper in the final set.
We agreed upon data to extract, including data source, research questions, main findings, and design implications.
We independently read and extracted information from each article,
then compared values, discussed disagreements, and reached consensus.

Next, we independently grouped similar papers into affinity groups~\cite{beyer99} based on each study's primary focus.
We then compared groupings and discussed our rationale,
shifting several papers between groups based on this discussion.
Once we agreed upon the groups, we labeled each based on study focus (see Table~\ref{tab:lit_review}). 
We describe the results of this analysis in Section~\ref{lit_review}.

\subsection{Design Feature Identification} 
In order to identify the design decisions that shape CCS, two authors analyzed four real-world CCS implementations: Twitter/X's Community Notes\footnote{\url{https://x.com/i/communitynotes/}}, Meta's Community Notes\footnote{\url{https://transparency.meta.com/features/community-notes}} (which covers Facebook, Instagram, and Threads), TikTok's Footnotes\footnote{\url{https://support.TikTok.com/en/using-TikTok/exploring-videos/footnotes}}, and YouTube's Notes\footnote{\url{https://support.google.com/youtube/answer/14925346?hl=en}}. 
Our analysis was primarily based on public, company-written documentation and communications about the systems and their features, retrieved in July 2025, which we collectively refer to as \textit{articles}.
While we attempted to become contributors for all platforms, some (TikTok, YouTube) did not grant us access.
For those that did, this access complemented our understanding and supported our analysis.

\begin{table*}[]
    \centering
    \small 
    \begin{tabularx}{\textwidth}{p{3.18cm}p{5.1cm}X}
        \toprule

        \textbf{Study Focus} & \textbf{Primary Method} & \textbf{Papers} \\
        \midrule

        \raggedright System Impact (n=24) & \raggedright Data Analysis (X~Community Notes Data) & 
        \cite{bouchaudAlgorithmicResolutionCrowdsourced2025b,renaultCollaborativelyAddingContext2024a,chuai25,truong2025communitynotesvulnerablerater,slaughter25,chujyo25,chuaiCommunitybasedFactcheckingReduces2024a,rollout21,kimDifferentialImpactIndividual2025e,diffusion21,chuai2025requestnoterequestfunction,matamoros25,razuvayevskaya2025timelinessconsensuscompositioncrowd,kangur2024checkscheckersexploringsource,DeepDiveXs,gao24,jama24,bobek2025communityfactchecksbreakfollower,borwankarDemocratizationMisinformationMonitoring2022a,Wirtschafter_Majumder_2023,wangEfficiencyCommunityBasedContent2024b,colemanLimitingFactorsEffectiveness2023,dekeulenaarTwitterDemotionCommunity2025,cham}
        \\
        \addlinespace
        
        Contributor Behavior (n=16) & \raggedright Data Analysis (X~Community Notes Data) &
        \cite{allen22,borenstein2025,pilarskiCommunityNotesVs2023,pröllochs2021communitybasedfactcheckingtwittersbirdwatch,cikm22,kuuse25,jones22,solovevReferencesUnbiasedSources2025a,arjmandilari2025threatssustainabilitycommunitynotes,toyoda25,Phillips2025,renault25,fraxanetUnpackingPolarizationAntagonism2024c,martelPoliticalMotivesHelp2025,mohammadi2025birdwatchcommunitynotestwitter,yoon25}
        \\
        \addlinespace
        
        \raggedright System Design (n=10) & User studies, design specs, algorithm analysis & 
        \cite{mohammadiAIFeedbackEnhances2025a,wu2025crowdllmaugmentedcommunitynotes,wojcikBirdwatchCrowdWisdom2022b,hawkeye,franzmeyer2024hellofreshllmevaluationsstreams,liScalingHumanJudgment2025a,supernotes,zhang2025commenotessynthesizingorganiccomments,singh2025on,xing2025communitynotesdatasetexploringhelpfulness}\\
        \addlinespace

        \raggedright Users' Perceptions (n=6) & User studies & \cite{sharevski2025helps,liu23,shusas24,drolsbachCommunityNotesIncrease2024a,Kankham18062025,Hameleers18052024}\\
    \bottomrule
    \end{tabularx}
    \caption{The n=56 articles found in our literature review, grouped by study focus.}
    \label{tab:lit_review}
    \Description{Summary of existing research categorized by study focus, showing four main categories: System Impact (10 studies using data analysis of X Community Notes data), Contributor Behavior (9 studies using data analysis), User's Perceptions (6 studies using user studies), and System Design (5 studies using various methods including user studies and algorithm analysis). Each row includes sample paper titles.}
\end{table*}

For each platform, we independently read the related articles and extracted design details into a spreadsheet to produce a set of design features.
We then convened and performed a round of affinity diagramming~\cite{beyer99}, in which we wrote the design features on sticky notes and placed them on a whiteboard.
We placed similar design features together, and through this inductive method identified feature groups.
We then used thematic analysis~\cite{braunUsingThematicAnalysis2006} to identify the themes of the groupings and came up with provisional group names.
We repeated this process with each platform, reconvening to do another round of affinity diagramming after each round of feature extraction.
Finally, we synthesized these design features and feature groups into the aspects, dimensions, and options of the design space, which we present below in Section~\ref{design-space}.

\section{What Does the Research Community Know About Twitter-Style CCS?}\label{lit_review}

Our systematic literature review of Twitter-style CCS research found n=56 papers and identified four high-level categories of study: studies of the system's impact, of contributor behavior, of alternative system designs, and of users' perceptions of crowdsourced context.
These categories naturally overlap and are not mutually exclusive, but we highlight them to emphasize the different types of knowledge emerging from this body of research.
Table~\ref{tab:lit_review} groups the papers according to their most significant contribution, though several papers make multiple contributions and thus span categories.

The most common study type explored CCS impact on the implementing platform (n=24 papers). 
These studies all analyzed data from X's Community Notes, asking both causal questions---such as the effect of displaying notes on post diffusion~\cite{diffusion21,rollout21}---and descriptive questions about how X's CCS performed in specific settings, such as during natural disasters~\cite{chujyo25,bouchaudAlgorithmicResolutionCrowdsourced2025b}.
Studies found evidence that CCS can reduce both the spread~\cite{diffusion21} and believability~\cite{wojcikBirdwatchCrowdWisdom2022b} of misleading posts annotated with notes.
Several articles examined the bridging algorithm used in X's Community Notes, which is designed to identify high-quality crowdsourced notes likely to be perceived as helpful by users across an ideological spectrum~\cite{bouchaudAlgorithmicResolutionCrowdsourced2025b}.
Studies showed that while bridging can be effective~\cite{bouchaudAlgorithmicResolutionCrowdsourced2025b}, this approach takes significant time to identify ``helpful'' notes, often causing substantial delays before notes are displayed~\cite{renaultCollaborativelyAddingContext2024a,chuaiCommunitybasedFactcheckingReduces2024a,rollout21}. 
Further, studies have shown that the vast majority of suggested notes ($\sim90\%$) are never published due to insufficient ratings or lack of consensus ~\cite{chujyo25,Wirtschafter_Majumder_2023}.

Another common study type (n=16) examined CCS contributor behavior.
While one study used interviews~\cite{yoon25}, all others analyzed open data from X's Community Notes. 
Studies asked descriptive questions about contributor actions, such as what types of posts~\cite{pilarskiCommunityNotesVs2023,cikm22} and accounts~\cite{allen22} they add context to, what content they include in notes~\cite{jones22,kuuse25}, and what types of notes they rate as most helpful~\cite{solovevReferencesUnbiasedSources2025a,Phillips2025}. 
Three key findings emerged: 1) contributors show political bias in post selection and note ratings~\cite{allen22}; 
2) contributors behave differently than professional fact-checkers---for example, by focusing on larger accounts~\cite{pilarskiCommunityNotesVs2023}--- yet often cite them as sources~\cite{borenstein2025};
and 3) contributors perceive certain note types as more helpful than others. 
For example, notes that support claims with links to unbiased external sources (i.e., sources assessed as politically neutral by a third-party organization\footnote{\url{https://mediabiasfactcheck.com/}}) ~\cite{solovevReferencesUnbiasedSources2025a} are rated more helpful than those that do not.

A third study type was design studies that either proposed or built CCS variations (n=10). 
These papers focused on ways to improve the quantity~\cite{liScalingHumanJudgment2025a} and quality~\cite{mohammadiAIFeedbackEnhances2025a} of notes, as well as techniques to improve the performance of the note curation algorithm~\cite{hawkeye}.
The bridging algorithm for selecting ``helpful'' notes was introduced in one of these papers~\cite{wojcikBirdwatchCrowdWisdom2022b}, as a solution to bias~\cite{allen22} in a majority-rule note selection system. 
Recent papers have proposed ways of incorporating AI into the note-writing process and found that LLMs have the potential to generate helpful notes~\cite{supernotes,mohammadiAIFeedbackEnhances2025a}.

A final study type sought to understand users' (not CCS contributors as above) perceptions of the notes published by CCS (n=6).
These studies used experiments~\cite{liu23}, interviews~\cite{shusas24}, and analysis of X's data~\cite{Hameleers18052024} to answer descriptive questions about how individuals perceived the credibility and helpfulness of notes published by CCS.
These studies found evidence that users perceive CCS notes as just as, or even more, credible than context from professional fact checkers~\cite{liu23,shusas24}.

While these studies begin to explore X's CCS implementation and its impact, much remains unexplored.
As we show next, the CCS design space is potentially much wider than demonstrated by existing research.

\section{A Design Space for CCS}\label{design-space}

\begin{table*}[]
\centering
\small
\begin{tabular}{p{9cm}>{\centering}p{1.1cm}>{\centering}p{.8cm}>{\centering}p{1.2cm}>{\centering\arraybackslash}p{1.2cm}}
\toprule
\textbf{Dimension} & \textbf{TikTok} & \textbf{Meta} & \textbf{Twitter/X} & \textbf{YouTube} \\
\midrule

\multicolumn{5}{l}{\textit{Who sees published notes?}} \\
All users. & & \checkmark & \checkmark & \\
Only users from a specific geo/market. & \checkmark & & & \checkmark \\
Only users that have opted in. & & & & \\
\midrule

\multicolumn{5}{l}{\textit{Who are the context contributors?}} \\
Users. & \checkmark & \checkmark & \checkmark & \checkmark \\
Platform employees/contractors. & & & & \\
Third parties. & & & & \\
AI systems. & & & \checkmark & \\
\midrule

\multicolumn{5}{l}{\textit{Who are the rating contributors?}} \\
Users. & \checkmark & \checkmark & \checkmark & \\
Platform employees/contractors. & & & & \checkmark \\
Third parties. & & & & \\
AI Systems. & & & & \\
\midrule

\multicolumn{5}{l}{\textit{What must a user do to become a context contributor?}} \\
Nothing: all users are contributors by default. & & & & \\
Nothing: users cannot be context contributors. & & & & \\
Apply. & \checkmark & \checkmark & \checkmark & \checkmark \\
Receive an invite from the platform. & & & & \checkmark \\
Be admitted by the platform. & \checkmark & \checkmark & \checkmark & \\
Meet a quality threshold of past performance. & & & \checkmark & \\
\midrule

\multicolumn{5}{l}{\textit{What must a user do to become a rating contributor?}} \\
Nothing: all users are contributors by default. & & & & \\
Nothing: users cannot be rating contributors. & & & &  \checkmark \\
Apply. & \checkmark & \checkmark & \checkmark & \\
Receive an invite from the platform. & & & & \\
Be admitted by the platform. & \checkmark & \checkmark & \checkmark & \\
Meet a quality threshold of past performance. & & & & \\
\midrule

\multicolumn{5}{l}{\textit{What types of eligibility requirements limit who can be a contributor?}} \\
User-age requirements. & \checkmark & \checkmark & & \\
Account-age requirements. & \checkmark & \checkmark & \checkmark & \checkmark \\
Lack of policy-violations. & \checkmark & \checkmark & \checkmark & \checkmark \\
Geographic requirements. & \checkmark & \checkmark & \checkmark & \\
Language requirements. & & \checkmark & & \checkmark \\
Account authentication requirements. & \checkmark & \checkmark & \checkmark & \\
Proof of some real-world credentials. & & & & \\
\midrule

\multicolumn{5}{l}{\textit{Is the number of contributors limited?}} \\
Yes: platform has discretion over who they admit. & \checkmark & \checkmark & \checkmark & \checkmark \\
No: all who meet eligibility criteria are automatically admitted. & & & & \\
\midrule

\multicolumn{5}{l}{\textit{How are contributors incentivized?}} \\
Paid by the platform. & & & & \checkmark \\
Pseudonymous reputation. & & & \checkmark & \\
Account reputation. & & & & \\
Unclear: contributor behavior is not visible. &  \checkmark &  \checkmark & & \\
\midrule

\multicolumn{5}{l}{\textit{How are notes and ratings solicited?}} \\
Contributors add notes to posts they organically see on the platform. & \checkmark & \checkmark & \checkmark & \checkmark \\
Contributors have feeds of posts that need notes, and notes that need ratings. & & \checkmark & \checkmark & \\
Platform notifies contributors of specific posts and notes to review. &  &  \checkmark & \checkmark & \\
Users can request notes on posts or additional review on published notes. & & \checkmark & \checkmark & \checkmark \\

\bottomrule
\end{tabular}
\caption{Participation dimensions and options, as well as platforms that documented these design choices.}
\label{tab:participation_dimensions}
\Description{Participation dimensions and options for CCS design, showing 9 dimensions including who sees published notes, who can be contributors, eligibility requirements, contributor limits, incentives, and solicitation methods. Each dimension lists multiple implementation options with corresponding platforms that have adopted each approach}
\end{table*}

Our analysis of the various CCS implementations revealed common characteristics shared between the systems as well as places for variation.
Analyzing CCS features allows us to construct a \textit{design space} of possibilities for these systems.
Within this design space, we identify six key \textit{aspects} of CCS:
\begin{enumerate}
    \item \textit{Participation}: Who can use the system and what are different users allowed to do?
    \item \textit{Inputs}: What form do contributions to the system take?
    \item \textit{Curation}: How are suggested notes selected for display?
    \item \textit{Presentation}: How are selected notes displayed?
    \item \textit{Platform Treatment}: How does the platform treat posts with displayed notes?
    \item \textit{Transparency}: How may the public observe and evaluate the system's functioning?
\end{enumerate}
Within each aspect, we identify several \textit{dimensions} (i.e., points of potential variation), which we frame as design questions, and \textit{options}
(i.e., potential answers for each dimension) based on the features observed in our analysis.
In the tables in the sections that follow, we indicate when we found evidence that a platform implemented a certain option.
However, platforms did not document all of their design choices, so our representations of each platforms' features should be seen as illustrative, not definitive.
In constructing the design space, we additionally include several speculative options not seen in current implementations, but which demonstrate important potential variations.
The sections that follow detail the dimensions and options for each aspect.

\subsection{Participation}

We define \textit{Participation} as the design choices that shape \textit{who} uses the system and \textit{what} different users are allowed to do.
Table~\ref{tab:participation_dimensions} shows the participation dimensions and options derived from our analysis.
Platform choices in this aspect range from those that encourage user participation in the system to those that discourage it.

The most basic dimension is, \textit{``Who sees published Notes?''}
Platforms' answers to this question greatly affect the system's impact.
The most participatory implementations, like X's, allow all users to see these notes, while less participatory implementations limit the audience, for example by location. 
An even less participatory option is making notes an optional feature that users must opt into, which would greatly limit system impact.

Other dimensions of participation determine who can \textit{contribute} to the system, such as: \textit{``Who are the context contributors?''}
These dimensions shape the quality, quantity, and variety of the notes the system produces. 
All implementations we examined rely on platform users as contributors, but we see variation in \textit{what} these users are allowed to contribute.
For example, YouTube does not allow users to provide ratings; instead, ratings are provided by third parties paid by the platform. 
X allows AI systems, in addition to users, to contribute notes via an API.
All implementations impose eligibility restrictions on users, limiting who can be a contributor.
Yet platforms also decide which eligible applicants are admitted.
A more participatory system would automatically admit all eligible applicants, or could eliminate eligibility restrictions and make all users contributors by default.

Beyond decisions that limit the set of possible contributors, platforms must also consider how to encourage contributor  participation: \textit{``How are contributors incentivized?''}
While platforms could pay contributors,
all implementations we examined rely on volunteer labor to some extent.
These approaches may incentivize contributions through reputation systems, for example by publicly associating contributors with their user accounts or with pseudonyms.
Another important question is: \textit{``How are notes and ratings solicited?''} 
All platforms allow contributors to add notes to posts they organically encounter.
However, this design choice may bias which posts receive notes if, for example, contributors only follow accounts from one end of the ideological spectrum. 
Section~\ref{curation} explores ways to reduce bias when selecting which notes to publish, but the choices here affect the \textit{supply} of notes available for selection.
Some platforms take more active approaches and use in-app feeds or notifications to solicit notes for specific posts or ratings for specific suggested notes, as shown in Table~\ref{tab:participation_dimensions}. 
These approaches can counter supply-side bias or encourage notes on posts most likely to cause harm, such as those going viral. 

\subsection{Inputs}

\begin{table*}[]
\centering
\small
\begin{tabular}{p{9cm}>{\centering}p{1.1cm}>{\centering}p{.8cm}>{\centering}p{1.2cm}>{\centering\arraybackslash}p{1.2cm}}
\toprule
\textbf{Dimension} & \textbf{TikTok} & \textbf{Meta} & \textbf{Twitter/X} & \textbf{YouTube} \\
\midrule

\multicolumn{5}{l}{\textit{What tools are note contributors given?}} \\
Free-text inputs. & \checkmark & \checkmark & \checkmark & \checkmark \\
Ability to add source links. & & \checkmark & & \checkmark \\
Ability to add references to a portion of the post (e.g. timestamp in the video being noted). & & & & \checkmark \\
AI note writing assistance. & & & & \\
API for AI note creation. & & & \checkmark & \\
\midrule

\multicolumn{5}{l}{\textit{What tools are rating contributors given?}} \\
A fixed set of rating options (i.e., helpful, somewhat helpful, not helpful). & \checkmark & \checkmark & \checkmark & \checkmark \\
A free-text explanation field. & & & & \checkmark \\
A fixed set of explanation options. & & \checkmark & \checkmark & \\
API for AI rating creation. & & & & \\
\midrule

\multicolumn{5}{l}{\textit{Which posts can have notes left on them?}} \\
All. & & & \checkmark & \\
Posts by specific account types (public, verified, pages). & \checkmark & \checkmark & & \\
Posts of specific types of content (no ads, none that may feature minors). & & & & \checkmark \\
Posts that reach a certain level of engagement. & & & & \\
Posts that have been flagged by users as needing additional context. & & & & \\
\midrule

\multicolumn{5}{l}{\textit{What restrictions apply to notes?}} \\
Must adhere to community guidelines. & \checkmark & \checkmark & \checkmark & \checkmark \\
Length restrictions. & & \checkmark & & \checkmark \\
Cannot be edited once submitted. & & \checkmark & \checkmark & \\
Cannot be deleted once submitted. & & & & \\
Source URL required. & & \checkmark & & \\
\midrule

\multicolumn{5}{l}{\textit{What restrictions apply to ratings?}} \\
Must adhere to community guidelines. & & & & \checkmark \\
Length restrictions. & & & & \checkmark \\
Cannot be edited once submitted. & & & \checkmark & \checkmark \\
Cannot be deleted once submitted. & & & \checkmark & \checkmark \\
\midrule

\multicolumn{5}{l}{\textit{Can the published note on a post change over time?}} \\
Published notes can be hidden by the crowd indefinitely. & \checkmark & \checkmark & & \checkmark \\
Note authors can edit or delete their notes. & & \checkmark & & \checkmark \\
Post authors can edit/remove notes on their posts. & & & & \\
Published notes are "locked" after a certain amount of time. & & & \checkmark & \\

\bottomrule
\end{tabular}
\caption{Inputs dimensions and options, as well as platforms that documented these design choices.}
\label{tab:inputs_dimensions}
\Description{Inputs dimensions covering what tools are provided to contributors, which posts can have notes, restrictions on notes and ratings, and whether notes can change over time. Shows the various input methods and constraints across different platforms.}
\end{table*}

We define \textit{Inputs} as the design choices that shape the type and content of contributions to the system. 
Table~\ref{tab:inputs_dimensions} shows the associated dimensions and options for the inputs dimension.

Contributions are shaped by the affordances that platforms provide contributors, including the user interfaces for producing notes and ratings. 
These interfaces may allow free-form or structured input. 
Such differences may affect what information contributions contain and how they appear.
For example, Meta's implementation has a required ``source URL'' field, which, based on findings from Section~\ref{lit_review}, may lead to more helpful notes.
Contributions are also shaped by additional platform restrictions.
For example, the implementations we examined required that contributions adhere to the platforms' content policies.
Finally, design choices that give contributors fixed options may both be easier to use and produce more consistent contributions. 
In these ways, design decisions in this aspect affect the volume and quality of both notes and ratings.

A final interesting question is, \textit{``Can the published note on a post change over time?''}
Some implementations allow contributors to edit or delete their notes, while others do not.
Additionally, the published note on a post can change if the curation process that selects notes (discussed below) is re-run over time.
To deal with this, X's implementation has a feature that ``locks'' the published note after a certain amount of time, preventing it from changing.

\subsection{Curation}\label{curation}

\begin{table*}[]
\centering
\small
\begin{tabular}{p{9cm}>{\centering}p{1.1cm}>{\centering}p{.8cm}>{\centering}p{1.2cm}>{\centering\arraybackslash}p{1.2cm}}
\toprule
\textbf{Dimension} & \textbf{TikTok} & \textbf{Meta} & \textbf{Twitter/X} & \textbf{YouTube} \\
\midrule

\multicolumn{5}{l}{\textit{What determines which notes are shown alongside posts?}} \\
Algorithmic decision based on data from contributors. & \checkmark & \checkmark & \checkmark & \checkmark \\
Editorial decision made by the platform. & & & & \\
Editorial decision made by contributors. & & & & \\
\midrule

\multicolumn{5}{l}{\textit{What (alleged) control does the platform have over what is displayed?}} \\
Platform chooses what is or is not published. & & & & \\
Platform can edit note content. & & & & \\
Platform can remove notes if they violate platform content policy. & \checkmark & \checkmark & \checkmark & \checkmark \\
Platform has no control. & & & & \\
\midrule

\multicolumn{5}{l}{\textit{What data is used to make an algorithmic decision?}} \\
Ratings. & \checkmark & \checkmark & \checkmark & \checkmark \\
Note content (length, inclusion of a link, etc). & & & & \\
A note author's past platform behavior as a contributor. & & & & \\
A note author's past platform behavior as a regular user. & & & & \\
Demographic/personal information about the note author. & & & & \\
Rating contributor's past platform behavior as a contributor. & \checkmark & \checkmark & \checkmark & \checkmark \\
Rating contributor's past platform behavior as a regular user. & & & & \\
Demographic/personal information about the rating contributor. & & & & \\
\midrule

\multicolumn{5}{l}{\textit{What is the aim of the algorithmic decision?}} \\
Identify notes that appeal across the ideological spectrum (``bridging''). & \checkmark & \checkmark & \checkmark & \checkmark \\
Identify notes with majority appeal. & & & & \\
Identify factually-accurate notes. & & & & \\
Identify notes that reference reliable sources. & & & & \\

\bottomrule
\end{tabular}
\caption{Curation dimensions and options, as well as platforms that documented these design choices.}
\label{tab:curation_dimensions}
\Description{Curation dimensions focusing on how notes are selected for display, including algorithmic decision processes, platform control levels, data sources used, and aims of algorithmic decisions. Primarily shows bridging-based approaches across platforms.}
\end{table*}

We define \textit{Curation} as the design choices that affect how notes are selected for display.
Table~\ref{tab:curation_dimensions} shows all of the associated dimensions and options.
Design decisions here affect the quality of the displayed notes, the potential for bias in note selection, as well as the speed and volume at which notes appear.

Perhaps the most important design question is \textit{``What process determines which notes are shown alongside posts?''}
The implementations that we studied all use bridging algorithms to, broadly speaking, identify notes that have been rated as ``helpful'' by contributors who have disagreed in their past ratings.
While these bridging algorithms are currently popular, alternatives are possible: for example, Twitter's first implementation of Birdwatch used a majority voting algorithm. 
Alternatively, bridging goals could be achieved using contributor information other than their past ratings, such as the accounts that they follow. 
Algorithmic approaches are also not the only option, as CCS could select notes via editorial decisions by either the platform or contributors.
Another possible approach is selecting notes via digital juries~\cite{fanDigitalJuriesCivicsOriented2020} of users, who could be selected either randomly or based on other criteria.

A related question is: \textit{``What (alleged) control does the platform have over what is displayed?''}
Platforms can exert varying levels of control over system output, although current implementations all intentionally minimize the platform's role. 
Still, platforms retain control over editorial decisions in cases of content policy violations.

\subsection{Presentation}

\begin{table*}[]
\centering
\small
\begin{tabular}{p{9cm}>{\centering}p{1.1cm}>{\centering}p{.8cm}>{\centering}p{1.2cm}>{\centering\arraybackslash}p{1.2cm}}
\toprule
\textbf{Dimension} & \textbf{TikTok} & \textbf{Meta} & \textbf{Twitter/X} & \textbf{YouTube} \\
\midrule

\multicolumn{5}{l}{\textit{Where are published notes presented to users?}} \\
Below the text of the original post. & \checkmark & \checkmark & \checkmark & \checkmark \\
Above the text of the original post. & & & & \\
In an overlay on top of the original post. & & & & \\
On both the original post and any re-shares of it. & & & \checkmark & \\
\midrule

\multicolumn{5}{l}{\textit{Who is notified about notes when they are published?}} \\
Original post authors. & \checkmark & \checkmark & \checkmark & \checkmark \\
Contributors who have written/rated the published note. & & \checkmark & & \checkmark \\
Every user that replied, liked, or reposted the original post. & & \checkmark & \checkmark & \\
Every user that was exposed to the original post. & & & & \\
\midrule

\multicolumn{5}{l}{\textit{How many notes are shown on a post?}} \\
At most one. & \checkmark & \checkmark & \checkmark & \checkmark \\
Multiple. & & & & \\
\midrule

\multicolumn{5}{l}{\textit{How is the author of a note shown on a post?}} \\
The author of the note is anonymous/omitted completely. & \checkmark & \checkmark & \checkmark & \checkmark \\
% The author of the \textit{note} is pseudonymous. & & & \checkmark & \\
The actual identity of the note author is shown. & & & & \\
\midrule

\multicolumn{5}{l}{\textit{What information about a note's ratings are shown on a post?}} \\
Only whether the note is rated ``helpful''. & \checkmark & \checkmark & \checkmark & \checkmark \\
Some indicator of how helpful the note is. & & & & \\

\bottomrule
\end{tabular}
\caption{Presentation dimensions and options, as well as platforms that documented these design choices.}
\label{tab:Presentation_dimensions}
\Description{Presentation dimensions covering where notes are displayed, who gets notified, how many notes are shown, and how authorship and ratings are presented. Most platforms show notes below posts with anonymous authorship.}
\end{table*}

We define \textit{Presentation} as the design choices affecting how published notes are displayed to users.
Table~\ref{tab:Presentation_dimensions} shows all associated dimensions and options.
Design decisions here affect the impact of published notes on users who see the associated posts.

The biggest question is: \textit{``Where are published notes presented to users?''}
Answers to this question affect whether users will ever see published notes.
Note presentation can range from aggressive to subtle: notes can be mandatory overlays that users must dismiss to see the associated post; notes can be shown alongside posts where they can be easily ignored; or platforms may simply indicate that a note exists and require users to click to view it.
Additionally, there is the question of which instances of the post display the published note: just the original, or also re-shares that appear across the platform.
This affects the number of users who see a published note.
Platforms must also decide: \textit{``Who is notified about notes when they are published?''}
Notifications can retroactively influence the understanding of users who saw a post \textit{before} it had a published note. 

Other important questions affect the amount of information about notes conveyed to users. 
Twitter's original Birdwatch implementation showed several ``helpful'' notes at a time, but all implementations we examined show only one.
Relatedly, platforms may provide details about a note's author or ratings, which could inform how users interpret  notes. 

\subsection{Platform Treatment}

\begin{table*}[]
\centering
\small
\begin{tabular}{p{9cm}>{\centering}p{1.1cm}>{\centering}p{.8cm}>{\centering}p{1.2cm}>{\centering\arraybackslash}p{1.2cm}}
\toprule
\textbf{Dimension} & \textbf{TikTok} & \textbf{Meta} & \textbf{Twitter/X} & \textbf{YouTube} \\
\midrule

\multicolumn{5}{l}{\textit{How are posts with notes treated by the platform?}} \\
No different from any other posts. & \checkmark & \checkmark & \checkmark & \checkmark \\
Limit monetization options. & & & & \\
Limit positioning next to ads. & & \checkmark & & \\
Limit spread (disabling reshares, deranking in feeds) & & & & \\
Increase spread (boost ranking in feeds) & & & & \\
\midrule

\multicolumn{5}{l}{\textit{How do notes interact with other content moderation features?}} \\
Replaces third-party fact checking program. & & \checkmark & \checkmark & \\
Complements platform-provided content moderation. & \checkmark & \checkmark & \checkmark & \checkmark \\
Notes are presented on posts alongside other ``information reliability'' indicators. & \checkmark & & & \checkmark \\

\bottomrule
\end{tabular}
\caption{Platform Treatment dimensions and options, as well as platforms that documented these design choices.}
\label{tab:treatment_dimensions}
\Description{Platform Treatment dimensions showing how posts with notes are treated by platforms, including monetization restrictions, interaction with other moderation features, and relationship to fact-checking programs.}
\end{table*}

We define \textit{Platform Treatment} as the design choices about how the platform treats posts that have published notes.
Table~\ref{tab:treatment_dimensions} shows all associated dimensions and options.
Design decisions here affect the impact of CCS on post spread.

The central question is: \textit{``How are posts with notes treated by the platform?''}
All current implementations emphasize that they do not algorithmically penalize posts with notes (for example, by down-ranking them in recommendation algorithms).
But this need not be the case: perhaps certain note types provide a signal about low post quality and platforms may want to suppress associated posts.
Alternatively, platform may impose restrictions on posts with notes, such as prohibiting monetization or display next to ads.

It is also important to ask: \textit{``How do notes interact with other content moderation features?''}
Some platforms have shut down their fact-checking programs and replaced them with CCS.
However, all platforms we surveyed still have content moderation systems of some kind, which CCS complement.
Finally, YouTube and TikTok display multiple ``information reliability'' indicators alongside posts in tandem with notes, such as platform-provided banners or labels. 

\subsection{Transparency}

\begin{table*}[]
\centering
\small
\begin{tabular}{p{9cm}>{\centering}p{1.1cm}>{\centering}p{.8cm}>{\centering}p{1.2cm}>{\centering\arraybackslash}p{1.2cm}}
\toprule
\textbf{Dimension} & \textbf{TikTok} & \textbf{Meta} & \textbf{Twitter/X} & \textbf{YouTube} \\
\midrule

\multicolumn{5}{l}{\textit{How public is the note curation system?}} \\
Open source. & & & \checkmark & \\
Open data. & & & \checkmark & \\
Fully verifiable. & & & & \\
Not public. & \checkmark & \checkmark & & \checkmark \\
\midrule

\multicolumn{5}{l}{\textit{How public is the contributor admission process?}} \\
Open source. & & & & \\
Open data. & & & & \\
Fully verifiable. & & & & \\
Not public. & \checkmark & \checkmark & \checkmark & \checkmark \\
\midrule

\multicolumn{5}{l}{\textit{How public is the data needed to evaluate the impact of notes on engagement?}} \\
Open data. & & & & \\
Available to researchers. & & & & \\
Available to paying customers. & & & \checkmark & \\
Not public. & \checkmark & \checkmark & & \checkmark \\
\midrule

\multicolumn{5}{l}{\textit{What sort of documentation about the system is public?}} \\
Blog posts. & \checkmark & \checkmark & & \checkmark \\
Help articles/FAQs. & & \checkmark & & \\
Technical documentation. & & & \checkmark & \\
None. & & & & \\

\bottomrule
\end{tabular}
\caption{Transparency dimensions and options, as well as platforms that documented these design choices.}
\label{tab:transparency_dimensions}
\Description{Transparency dimensions covering public access to curation systems, contributor admission processes, engagement impact data, and documentation availability. Shows X/Twitter as most transparent with open source and open data approaches, while other platforms are mostly private.}
\end{table*}

We define \textit{Transparency} as the design choices that influence how the public can observe and evaluate system functioning.
Table~\ref{tab:transparency_dimensions} shows all associated dimensions and options.
Design decisions here may affect the perception of CCS by users, the public, and the research community. 

Of central interest is: \textit{``How public is the note curation system?''}
CCS source code and data can be public to allow outside study.
In addition to being public, source code and data can be made \textit{verifiable}, allowing outsiders to verify that the code and data indeed powers the platform's implementation.
Alternatively, platforms could keep everything private, as all platforms except X currently do.

To fully evaluate the efficacy of such a system, we also need to ask: \textit{``How public is the data needed to evaluate the impact of notes on engagement?''}
This data would allow the public to holistically evaluate CCS against other content moderation systems.
Currently, this data is very difficult to access, making holistic assessment of these systems challenging.

\section{Discussion}
Through our systematic literature review and analysis of real-world CCS implementations, we have produced a framework that can help the HCI research community better reason about, and design, these systems.
In this section, we discuss future directions for CCS research in HCI.
Specifically, we focus on how CCS fit within the broader field of content moderation, highlight underexplored portions of the design space, and reflect on the normative implications of CCS design choices.

\subsection{Can CCS Replace Content Moderation?}
CCS were created with the stated goal of addressing the misinformation problem~\cite{IntroducingBirdwatchCommunitybaseda}, but it is unclear if the platforms implementing them are serious about this aim.
For example, many see Meta's choice to replace their U.S. fact-checking program with a CCS~\cite{heatheraMoreSpeechFewer2025} as a thinly-veiled effort to curry political favor with a presidential administration critical of fact-checking~\cite{nixMetaEndsFactchecking2025}. 
In fact, from a platform's perspective, CCS have other appealing properties: they cost less than professional fact-checkers, they shield platforms from accusations of biased policy enforcement, and they offload the difficult responsibility of content moderation onto users.
It may be expedient for platforms to build CCS in the current moment, but it is unclear what the role of such systems will be if the political climate shifts.
Still, as long as CCS are in a position to exert influence over the information ecosystem, it is important for the HCI research community to study them.
Most importantly, we emphasize the need to better understand how CCS perform compared to other content moderation systems.
CCS may perform well in certain cases, but if on the whole they perform worse than the systems they are replacing, for example on the question of coverage, they should be used sparingly.

While it has been suggested that they can, CCS cannot effectively replace a fuller spectrum of content moderation tools, and were not designed to do so~\cite{elliottElonMusksMain}.
Content moderation is a broad term that encompasses a variety of different goals, including preventing crimes, mitigating lawful but awful behavior, and promoting high quality content~\cite{grimmelmann2015virtues}.
The tools used for content moderation are similarly varied and include both ``hard'' approaches, like content take-downs and account bans, and ``soft'' approaches, such as content labeling~\cite{zannettouWonElectionEmpirical2021}.
Systems that provide context are narrow in both their goal and mechanism: they are a soft moderation attempt to aid content interpretation by providing users with additional information~\cite{matamoros25}.
And yet, the empirical results summarized in this paper demonstrate that CCS may play a meaningful role in the content moderation ecosystem, and it is worth exploring how they may integrate with other parts of it. 
For example: how might CCS be integrated into platforms' \textit{automated} moderation systems? 
In the Platform Treatment design aspect, none of the implementations that we examined used the CCS output to affect the visibility of posts.
Perhaps these systems can be integrated into platform content recommendations algorithms to promote or demote posts based on their notes.

\subsection{Underexplored CCS Design Aspects}
Our design space reveals several ways forward for a HCI research agenda on CCS.
Several aspects are underexplored by current research, particularly the Participation aspect.
Encouraging positive participation in collaborative computing systems is a classic topic in HCI research~\cite{krautBuildingSuccessfulOnline2012} and one in which the community could make strong contributions.
One unexplored participation dimension is how platforms might more actively solicit notes and ratings from contributors. 
Past HCI work on how to best make use of the various skills of crowds may be instructive here~\cite{bernsteinSoylentWordProcessor2015,kitturFutureCrowdWork2013}.
It is worth noting that the more consequential these systems become, the more they will be targets for negative forms of participation as well, primarily gaming~\cite{danielQualityControlCrowdsourcing2018,epsteinWillCrowdGame2020}. 
Bridging algorithms are designed to be resistant to certain types of gaming~\cite{wojcikBirdwatchCrowdWisdom2022b}, but vulnerabilities may emerge.
One other area for study is the Presentation aspect, specifically how different approaches to visually displaying notes may affect user understanding of misleading claims~\cite{gamageLabelingSyntheticContent2025,martelMisinformationWarningLabels2023}.
Another promising area is the Inputs aspect, where recent work has begun to explore  the ways in which LLMs can supplement human effort and increase the quality and quantity of notes~\cite{supernotes,mohammadiAIFeedbackEnhances2025a}.
Of particular interest is X's recent decision to allow AI note contributors via a new API~\cite{liScalingHumanJudgment2025a}.
How such decisions affect the quantity and quality of notes should be a near-term focus of HCI research.
Of course, studying how changes to CCS impact information dynamics across platforms will be difficult without changes in the Transparency aspect.
Transparency about how CCS works (as X has provided to date) facilitates greater research insights.
Hopefully other platforms will soon adopt similar approaches to support ongoing studies of CCS effectiveness.

\subsection{Normative Guidelines for CCS Design}

CCS are complex sociotechnical systems~\cite{chernsPrinciplesSociotechnicalDesign1976} that integrate people, algorithms, and speech.
These systems' growing influence over the information shown on major social media platforms means that the design choices made in their implementation can have significant social consequences~\cite{friedmanBiasComputerSystems1996,flanaganEmbodyingValuesTechnology2008}.
With such high stakes, it is crucial that a human-centered research agenda consider the implications of different design choices, in order to holistically evaluate these systems' impacts.
One way to do so is to identify the primary normative concerns of such a system. 
We emphasize three interconnected concerns that should guide CCS design: \textit{user informedness}, \textit{distribution of power}, and \textit{fairness}.

\subsubsection{User Informedness}
While social media serves many different purposes, CCS are specifically concerned with social media's ability to inform the public.
Such systems should thus be evaluated by how well they facilitate user informedness: for example, the degree to which they inform users about the accuracy of information, minimize the spread of misleading posts, or otherwise improve information quality on a platform.
Our literature review surfaced ample evidence that the notes produced by CCS do have the power to increase user informedness, both by decreasing an individual's belief in misleading information, and by decreasing the spread of this information~\cite{wojcikBirdwatchCrowdWisdom2022b,renaultCollaborativelyAddingContext2024a,slaughter25,chuaiCommunitybasedFactcheckingReduces2024a,rollout21,diffusion21}.
Additionally, there is evidence~\cite{kimDifferentialImpactIndividual2025e} that these notes may produce less of a backfiring effect than social correction that is not mediated by a CCS~\cite{mosleh21}.
But while published notes have been shown to be effective at promoting understanding, we must also question how effective CCS are at producing these notes.
Several of these same studies found that Twitter's implementation of CCS produces notes too slowly to prevent the spread of viral content~\cite{rollout21,chuaiCommunitybasedFactcheckingReduces2024a,renaultCollaborativelyAddingContext2024a}.
There is also a significant gap in the research on the question of coverage---how many of the misleading posts on a platform receive published notes---which currently remains unanswered.
The research suggests that there is a CCS design trade-off between speed/coverage and quality/helpfulness: X's implementation has optimized for the latter, perhaps at the expense of the former.  
Whether this is the right decision should be up for debate.

For designers of CCS hoping to improve user informedness by increasing the system's speed and coverage, there are several aspects from our design space that could be relevant.
For example, within the Curation aspect, designers could explore bridging algorithm changes that could lead to faster identification of ``helpful'' notes.
Additionally, design changes within the Participation and Inputs aspects, such as those that would allow a larger contributor base or that would actively solicit notes on the most viral posts, could lead to a higher volume of notes and ratings, and thus quicker selection of helpful notes. 
When it comes to the difficulty of assessing the coverage of CCS, design changes within the Transparency aspect could make this easier to study, perhaps by providing easier access to post engagement metrics.

\subsubsection{Distribution of Power}
One of the selling points of CCS is that they shift power over content moderation from the platform to the community.
Giving users more of a say over the content that they see on their platform can make moderation decisions more legitimate and trusted~\cite{shusas24,liu23,drolsbachCommunityNotesIncrease2024a}.
Yet such claims require further scrutiny, and CCS should be evaluated to the extent that they actually empower users and do not merely allow platforms to shirk responsibility for misleading content.
Evidence from our literature review suggests that bridging algorithms, by design, fail to produce notes for the most polarizing content~\cite{bouchaudAlgorithmicResolutionCrowdsourced2025b}. 
For example, X's implementation of the bridging algorithm requires polarized contributors to agree that a note is helpful in order to publish it. 
The editorial function of the bridging algorithm can thus limit the type and scope of contributions in two important ways: in terms of framing, the algorithm may prefer centrist points of view and content that could be agreeable to both sides of an issue; in terms of selection, agreement on a note may not occur for contentious topics, preventing its publication.
Whether or not users at the ends of the ideological spectrum should be empowered to make moderation decisions is a valid question, but it is uncontroversial that the most polarizing content needs moderation, too.
Research has found evidence of harmful content being handled poorly by the system, either because potential notes never reach helpful status~\cite{chujyo25}, or because notes that do are not actually helpful for certain types of content, such as disinformation that is couched in humor~\cite{matamoros25}. 
Zooming out, the low publishing rate for suggested notes (around ~10\% on X~\cite{Wirtschafter_Majumder_2023} and ~6\% on Meta\footnote{\url{https://x.com/guyro/status/1965800300557857053}}) gives us pause that these systems are actually empowering note authors.
A related, underexplored area for research involves studying which note contributors are empowered by CCS.
This research area is difficult to study holistically, as these systems are pseudonymous, but some work has begun to explore it qualitatively~\cite{yoon25} and via partnerships with platforms~\cite{allen22}.
More research is needed to understand how CCS may empower users of certain demographics or political views over others.

Our design space offers several ways that designers can further empower users via CCS.
First and foremost is innovation in the Participation aspect that would increase the set and diversity of contributors.
The challenge for the platforms will be ensuing that quality remains as the contributor base increases, which is a tension that has been thoroughly studied in the crowdsourcing literature~\cite{danielQualityControlCrowdsourcing2018}, for example.
Innovation in the Inputs aspect may increase the quality of notes across the board, perhaps by giving note authors helpful feedback~\cite{mohammadiAIFeedbackEnhances2025a} that empowers more users to make successful contributions.
Another route to shift power to users is to disempower the platform.
Though CCS are often described as ways to reduce platforms' centrality in content moderation, platforms still exert power over the system in important ways, such as by deciding who is admitted as a contributor.
Platforms could further guarantee their neutrality through changes to the Transparency aspect, such as by making their contributor selection process public.
But considering that X's CCS implementation is the only even partially open system that we examined, a first step is for other platforms to follow X's lead and open source their note-curation algorithm and the crowdsourced data that they collect.

\subsubsection{Fairness}
CCS are often touted for their potential to implement content moderation in an unbiased way.
Yet these systems are not inherently free from bias.
Results from our literature review show bias in contributor behavior, in terms of the types of accounts that they leave notes on, as well as how they rate suggested notes~\cite{allen22}.
Bridging algorithms have helped mitigate bias in the note selection process, with the caveats noted above, but additional bias can still be introduced---for example, in the note creation process, or in the selection of contributors.
To give a concrete example: X's bridging algorithm finds agreement among contributors, \textit{not} the entire user base.
If contributors' beliefs are not representative of users' beliefs then the algorithm's outcomes may be a biased representation of the community, and not necessarily have broad appeal among users.
There is some evidence that so far X's implementation has not suffered from this problem, but it is worth considering.
Also, accusations of bias in top-down moderation are often due to the opacity of such approaches and may be mitigated by transparency efforts.
Only X's implementation has provided \textit{any} transparency, and yet the lack of transparency about the contributor base has made analyzing it for bias impossible.
Finally, the approach of displaying context rather than removing posts is arguably fairer in that it lessens the damage from false positives, by empowering users to come to their own conclusions about content's veracity~\cite{pengRescuingCounterspeechBridgingBased2024a}.
However, there is a trade-off: leaving up questionable content, even with helpful context, can potentially do harm to those targeted by the content.

Designers should consider several aspects from our design space that can influence the fairness properties of a CCS.
First off, all platforms have taken the approach that it is fairer not to remove misleading content. 
However, designers may explore other options from the Platform Treatment dimension to identify certain cases where the platform should take action on content flagged by the crowd as misleading.
Second, current implementations tend to display a single, definitive note, even if there may be several that are determined to be ``helpful''.
The Presentation aspect offers alternatives that could further the aim of allowing users to come to their own conclusions, such as by showing multiple notes when there are multiple with helpful ratings.
Additionally showing information about the note author could be helpful, as it could help users identify the credibility of the source.
At the same time, this feature may be at odds with contributor privacy, which current implementations have opted to protect.
Finally, innovations in the Curation aspect could explore other methods of selecting notes to display that may be regarded as fairer.
Current bridging algorithm implementations use past ratings to identify contributors with diverse perspectives: perhaps it would be fairer to instead use in-app behavior?
This could provide a fuller picture of a contributor's ideological leanings.
Encouraging more Transparency would also give assurances of fairness, both in how contributors are selected, and how notes are published.

\subsection{Limitations}
Our work here may have several limitations. Since there was no agreed upon name for CCS before our paper, we decided to focus our literature review on studies of Twitter-style CCS, and thus may have missed earlier work that would have meaningfully informed the design space.
Our analysis of platforms' implementations of CCS is primarily based on public documentation from a specific point in time and may not perfectly reflect the actual features of the deployed systems.

\section{Conclusion}
Crowdsourced Context Systems are increasingly prevalent in the content moderation strategies of prominent social media platforms.
To understand these systems, their impacts, and potential design improvements, we conducted a systematic literature review of CCS research and analyzed several major platforms' implementations. 
Despite recent trends towards the adoption of CCS, we emphasize that they are only one piece of a larger content moderation suite, and advise against using them as drop-in replacements for traditional fact-checking approaches.
The framework we produced can be used to reason about these systems, and consists of a theoretical model and a design space.
We additionally discuss some of the key normative concerns that should be considered when designing CCS.
Such systems have potential to address the challenge of misinformation, but to realize this potential, future design research from the HCI community must focus on ways to understand and improve these systems' performance.

\begin{acks}
This material is partially based upon work supported by the National Science Foundation under CISE Graduate Fellowship Grant No. 2313998 and SaTC Grant No. 2120651, as well as the John D. and Catherine T. MacArthur Foundation.
Any opinions, findings, and conclusions or recommendations expressed in this material are those of the authors and do not necessarily reflect the views of the John D. and Catherine T. MacArthur Foundation or the National Science Foundation.
Thank you to %Nick Judd,
James Grimmelmann and Kenny Peng for feedback on earlier versions of this work.
\end{acks}

\bibliographystyle{ACM-Reference-Format}
\bibliography{sample-base}

@String{Computing = "Computing" }

@String{Computer = "{IEEE} Computer" }

@String{Springer = "Springer-Verlag" }

@article{friedmanBiasComputerSystems1996,
	title = {Bias in computer systems},
	volume = {14},
	issn = {1046-8188},
	url = {https://dl.acm.org/doi/10.1145/230538.230561},
	doi = {10.1145/230538.230561},
	number = {3},
	urldate = {2025-09-06},
	journal = {ACM Trans. Inf. Syst.},
	author = {Friedman, Batya and Nissenbaum, Helen},
	month = jul,
	year = {1996},
	pages = {330--347},
}

@incollection{flanaganEmbodyingValuesTechnology2008,
	edition = {1},
	title = {Embodying {Values} in {Technology}: {Theory} and {Practice}},
	isbn = {978-0-521-85549-5 978-0-521-67161-3 978-0-511-49872-5},
	shorttitle = {Embodying {Values} in {Technology}},
	url = {https://www.cambridge.org/core/product/identifier/CBO9780511498725A023/type/book_part},
	urldate = {2025-09-06},
	booktitle = {Information {Technology} and {Moral} {Philosophy}},
	publisher = {Cambridge University Press},
	author = {Flanagan, Mary and Howe, Daniel C. and Nissenbaum, Helen},
	editor = {Van Den Hoven, Jeroen and Weckert, John},
	month = mar,
	year = {2008},
	doi = {10.1017/CBO9780511498725.017},
	pages = {322--353},
}

@misc{truong2025communitynotesvulnerablerater,
      title={Community Notes are Vulnerable to Rater Bias and Manipulation}, 
      author={Bao Tran Truong and Siqi Wu and Alessandro Flammini and Filippo Menczer and Alexander J. Stewart},
      year={2025},
      eprint={2511.02615},
      archivePrefix={arXiv},
      url={https://arxiv.org/abs/2511.02615}, 
}

@inproceedings{chuai25,
author = {Chuai, Yuwei and Sergeeva, Anastasia and Lenzini, Gabriele and Pr\"{o}llochs, Nicolas},
title = {Community Fact-Checks Trigger Moral Outrage in Replies to Misleading Posts on Social Media},
year = {2025},
isbn = {9798400713941},
publisher = {Association for Computing Machinery},
address = {New York, NY, USA},
url = {https://doi.org/10.1145/3706598.3713909},
doi = {10.1145/3706598.3713909},
booktitle = {Proceedings of the 2025 CHI Conference on Human Factors in Computing Systems},
articleno = {962},
numpages = {23},
location = {Yokohama, Japan},
series = {CHI '25}
}

@article{chernsPrinciplesSociotechnicalDesign1976,
	title = {The {Principles} of {Sociotechnical} {Design}},
	volume = {29},
	issn = {0018-7267},
	url = {https://doi.org/10.1177/001872677602900806},
	doi = {10.1177/001872677602900806},
	language = {EN},
	number = {8},
	urldate = {2025-09-06},
	journal = {Human Relations},
	author = {Cherns, Albert},
	month = aug,
	year = {1976},
	pages = {783--792},
}

@article{eckerPsychologicalDriversMisinformation2022,
	title = {The psychological drivers of misinformation belief and its resistance to correction},
	volume = {1},
	issn = {2731-0574},
	url = {https://www.nature.com/articles/s44159-021-00006-y},
	doi = {10.1038/s44159-021-00006-y},
	language = {en},
	number = {1},
	urldate = {2025-12-04},
	journal = {Nature Reviews Psychology},
	author = {Ecker, Ullrich K. H. and Lewandowsky, Stephan and Cook, John and Schmid, Philipp and Fazio, Lisa K. and Brashier, Nadia and Kendeou, Panayiota and Vraga, Emily K. and Amazeen, Michelle A.},
	month = jan,
	year = {2022},
	pages = {13--29},
}

@inproceedings{spiekerDiversePerspectivesCan2023,
author = {Thebault-Spieker, Jacob and Venkatagiri, Sukrit and Mine, Naomi and Luther, Kurt},
title = {Diverse Perspectives Can Mitigate Political Bias in Crowdsourced Content Moderation},
year = {2023},
isbn = {9798400701924},
publisher = {Association for Computing Machinery},
address = {New York, NY, USA},
url = {https://doi.org/10.1145/3593013.3594080},
doi = {10.1145/3593013.3594080},
booktitle = {Proceedings of the 2023 ACM Conference on Fairness, Accountability, and Transparency},
pages = {1280–1291},
numpages = {12},
location = {Chicago, IL, USA},
series = {FAccT '23}
}

@article{jiangReasoningPoliticalBias2020,
	title = {Reasoning about {Political} {Bias} in {Content} {Moderation}},
	volume = {34},
	copyright = {Copyright (c) 2020 Association for the Advancement of Artificial Intelligence},
	issn = {2374-3468},
	url = {https://ojs.aaai.org/index.php/AAAI/article/view/7117},
	doi = {10.1609/aaai.v34i09.7117},
	language = {en},
	number = {09},
	urldate = {2025-09-11},
	journal = {Proceedings of the AAAI Conference on Artificial Intelligence},
	author = {Jiang, Shan and Robertson, Ronald E. and Wilson, Christo},
	month = apr,
	year = {2020},
	pages = {13669--13672},
}

@book{gillespieCustodiansInternetPlatforms2018,
	title = {Custodians of the {Internet}: {Platforms}, {Content} {Moderation}, and the {Hidden} {Decisions} {That} {Shape} {Social} {Media}},
	isbn = {978-0-300-23502-9},
	shorttitle = {Custodians of the {Internet}},
	language = {en},
	publisher = {Yale University Press},
	author = {Gillespie, Tarleton},
	month = jun,
	year = {2018},
}

@article{jhaverHumanMachineCollaborationContent2019,
	title = {Human-{Machine} {Collaboration} for {Content} {Regulation}: {The} {Case} of {Reddit} {Automoderator}},
	volume = {26},
	issn = {1073-0516},
	shorttitle = {Human-{Machine} {Collaboration} for {Content} {Regulation}},
	url = {https://dl.acm.org/doi/10.1145/3338243},
	doi = {10.1145/3338243},
	number = {5},
	urldate = {2025-09-11},
	journal = {ACM Trans. Comput.-Hum. Interact.},
	author = {Jhaver, Shagun and Birman, Iris and Gilbert, Eric and Bruckman, Amy},
	month = jul,
	year = {2019},
	pages = {31:1--31:35},
}

@article{scheuermanFrameworkSeverityHarmful2021,
	title = {A {Framework} of {Severity} for {Harmful} {Content} {Online}},
	volume = {5},
	issn = {2573-0142},
	url = {https://dl.acm.org/doi/10.1145/3479512},
	doi = {10.1145/3479512},
	language = {en},
	number = {CSCW2},
	urldate = {2025-09-11},
	journal = {Proceedings of the ACM on Human-Computer Interaction},
	author = {Scheuerman, Morgan Klaus and Jiang, Jialun Aaron and Fiesler, Casey and Brubaker, Jed R.},
	month = oct,
	year = {2021},
	pages = {1--33},
}

@inproceedings{herderDigitalJunkfoodSocial2024,
author = {Herder, Eelco and Staring, Jouke},
title = {Digital Junkfood on Social Media: To Each Their Own Poison},
year = {2024},
isbn = {9798400705953},
publisher = {Association for Computing Machinery},
address = {New York, NY, USA},
url = {https://doi.org/10.1145/3648188.3678163},
doi = {10.1145/3648188.3678163},
booktitle = {Proceedings of the 35th ACM Conference on Hypertext and Social Media},
pages = {126–135},
numpages = {10},
location = {Poznan, Poland},
series = {HT '24}
}

@article{preeceEmpathicCommunitiesBalancing1999,
	title = {Empathic communities: balancing emotional and factual communication},
	volume = {12},
	issn = {0953-5438},
	shorttitle = {Empathic communities},
	url = {https://doi.org/10.1016/S0953-5438(98)00056-3},
	doi = {10.1016/S0953-5438(98)00056-3},
	number = {1},
	urldate = {2025-09-11},
	journal = {Interacting with Computers},
	author = {Preece, Jenny},
	month = sep,
	year = {1999},
	pages = {63--77},
}

@article{gorwa20,
	title = {Algorithmic content moderation: {Technical} and political challenges in the automation of platform governance},
	volume = {7},
	url = {https://doi.org/10.1177/2053951719897945},
	doi = {10.1177/2053951719897945},
	number = {1},
	journal = {Big Data \& Society},
	author = {Gorwa, Robert and Binns, Reuben and Katzenbach, Christian},
	year = {2020},
}

@inproceedings{chouakiWhatNewsPeople2024a,
author = {Chouaki, Salim and Chakraborty, Abhijnan and Goga, Oana and Zannettou, Savvas},
title = {What News Do People Get on Social Media? Analyzing Exposure and Consumption of News through Data Donations},
year = {2024},
isbn = {9798400701719},
publisher = {Association for Computing Machinery},
address = {New York, NY, USA},
url = {https://doi.org/10.1145/3589334.3645399},
doi = {10.1145/3589334.3645399},
booktitle = {Proceedings of the ACM Web Conference 2024},
pages = {2371–2382},
numpages = {12},
location = {Singapore, Singapore},
series = {WWW '24}
}

@misc{pengRescuingCounterspeechBridgingBased2024a,
	title = {Rescuing {Counterspeech}: {A} {Bridging}-{Based} {Approach} to {Combating} {Misinformation}},
	shorttitle = {Rescuing {Counterspeech}},
	url = {http://arxiv.org/abs/2410.12699},
	doi = {10.48550/arXiv.2410.12699},
	urldate = {2025-09-10},
	publisher = {arXiv},
	author = {Peng, Kenny and Grimmelmann, James},
	month = oct,
	year = {2024},
}

@article{smallPolisEscalarDeliberacion2021,
	title = {Polis: {Escalar} de la deliberación mediante el mapeo de espacios de opinión de alta dimensión},
	issn = {2254-4135, 1130-6149},
	shorttitle = {Polis},
	url = {https://www.e-revistes.uji.es/index.php/recerca/article/view/5516},
	doi = {10.6035/recerca.5516},
	language = {en},
	urldate = {2025-09-10},
	journal = {RECERCA. Revista de Pensament i Anàlisi},
	author = {Small, Christopher},
	month = jul,
	year = {2021},
}

@article{Phillips2025, title={Emotional language reduces belief in false claims}, volume={20}, DOI={10.1017/jdm.2025.10019}, journal={Judgment and Decision Making}, author={Phillips, Samantha C. and Wang, Sze Yuh Nina and Carley, Kathleen M. and Rand, David G. and Pennycook, Gordon}, year={2025}, pages={e43}
}

@inproceedings{toyoda25,
author = {Toyoda, Keisuke and Fukuma, Tomoki and Noda, Koki and Ichikawa, Yoshiharu and Kambe, Kyosuke and Masubuchi, Yu and Someda, Hiroshi and Toriumi, Fujio},
title = {Understanding and Mitigating Polarization in Community Notes: Factors and Strategies for Improved Consensus},
year = {2025},
isbn = {9798400713316},
publisher = {Association for Computing Machinery},
address = {New York, NY, USA},
url = {https://doi.org/10.1145/3701716.3717541},
doi = {10.1145/3701716.3717541},
booktitle = {Companion Proceedings of the ACM on Web Conference 2025},
pages = {2694–2698},
numpages = {5},
location = {Sydney NSW, Australia},
series = {WWW '25}
}

@misc{ovadyaBridgingSystemsOpen2023,
	title = {Bridging {Systems}: {Open} {Problems} for {Countering} {Destructive} {Divisiveness} across {Ranking}, {Recommenders}, and {Governance}},
	shorttitle = {Bridging {Systems}},
	url = {http://arxiv.org/abs/2301.09976},
	doi = {10.48550/arXiv.2301.09976},
	urldate = {2025-09-10},
	publisher = {arXiv},
	author = {Ovadya, Aviv and Thorburn, Luke},
	month = jul,
	year = {2023},
}

@inproceedings{cardDesignSpaceInput1990,
author = {Card, Stuart K. and Mackinlay, Jock D. and Robertson, George G.},
title = {The design space of input devices},
year = {1990},
isbn = {0201509326},
publisher = {Association for Computing Machinery},
address = {New York, NY, USA},
url = {https://doi.org/10.1145/97243.97263},
doi = {10.1145/97243.97263},
booktitle = {Proceedings of the SIGCHI Conference on Human Factors in Computing Systems},
pages = {117–124},
numpages = {8},
location = {Seattle, Washington, USA},
series = {CHI '90}
}

@inproceedings{leeDesignSpaceIntelligent2024,
author = {Lee, Mina and Gero, Katy Ilonka and Chung, John Joon Young and Shum, Simon Buckingham and Raheja, Vipul and Shen, Hua and Venugopalan, Subhashini and Wambsganss, Thiemo and Zhou, David and Alghamdi, Emad A. and August, Tal and Bhat, Avinash and Choksi, Madiha Zahrah and Dutta, Senjuti and Guo, Jin L.C. and Hoque, Md Naimul and Kim, Yewon and Knight, Simon and Neshaei, Seyed Parsa and Shibani, Antonette and Shrivastava, Disha and Shroff, Lila and Sergeyuk, Agnia and Stark, Jessi and Sterman, Sarah and Wang, Sitong and Bosselut, Antoine and Buschek, Daniel and Chang, Joseph Chee and Chen, Sherol and Kreminski, Max and Park, Joonsuk and Pea, Roy and Rho, Eugenia Ha Rim and Shen, Zejiang and Siangliulue, Pao},
title = {A Design Space for Intelligent and Interactive Writing Assistants},
year = {2024},
isbn = {9798400703300},
publisher = {Association for Computing Machinery},
address = {New York, NY, USA},
url = {https://doi.org/10.1145/3613904.3642697},
doi = {10.1145/3613904.3642697},
booktitle = {Proceedings of the 2024 CHI Conference on Human Factors in Computing Systems},
articleno = {1054},
numpages = {35},
location = {Honolulu, HI, USA},
series = {CHI '24}
}

@article{zhangFormFromDesignSpace2024a,
author = {Zhang, Amy X. and Bernstein, Michael S. and Karger, David R. and Ackerman, Mark S.},
title = {Form-From: A Design Space of Social Media Systems},
year = {2024},
issue_date = {April 2024},
publisher = {Association for Computing Machinery},
address = {New York, NY, USA},
volume = {8},
number = {CSCW1},
url = {https://doi.org/10.1145/3641006},
doi = {10.1145/3641006},
journal = {Proc. ACM Hum.-Comput. Interact.},
month = apr,
articleno = {167},
numpages = {47}
}

@inproceedings{kripleanSupportingReflectivePublic2012,
author = {Kriplean, Travis and Morgan, Jonathan and Freelon, Deen and Borning, Alan and Bennett, Lance},
title = {Supporting reflective public thought with considerit},
year = {2012},
isbn = {9781450310864},
publisher = {Association for Computing Machinery},
address = {New York, NY, USA},
url = {https://doi.org/10.1145/2145204.2145249},
doi = {10.1145/2145204.2145249},
booktitle = {Proceedings of the ACM 2012 Conference on Computer Supported Cooperative Work},
pages = {265–274},
numpages = {10},
location = {Seattle, Washington, USA},
series = {CSCW '12}
}

@article{zannettouWonElectionEmpirical2021,
	title = {"{I} {Won} the {Election}!": {An} {Empirical} {Analysis} of {Soft} {Moderation} {Interventions} on {Twitter}},
	volume = {15},
	copyright = {Copyright (c) 2021 Association for the Advancement of Artificial Intelligence},
	issn = {2334-0770},
	shorttitle = {"{I} {Won} the {Election}!"},
	url = {https://ojs.aaai.org/index.php/ICWSM/article/view/18110},
	doi = {10.1609/icwsm.v15i1.18110},
	language = {en},
	urldate = {2025-12-01},
	journal = {Proceedings of the International AAAI Conference on Web and Social Media},
	author = {Zannettou, Savvas},
	month = may,
	year = {2021},
	pages = {865--876},
}

@inproceedings{kripleanIntegratingOndemandFactchecking2014,
author = {Kriplean, Travis and Bonnar, Caitlin and Borning, Alan and Kinney, Bo and Gill, Brian},
title = {Integrating on-demand fact-checking with public dialogue},
year = {2014},
isbn = {9781450325400},
publisher = {Association for Computing Machinery},
address = {New York, NY, USA},
url = {https://doi.org/10.1145/2531602.2531677},
doi = {10.1145/2531602.2531677},
booktitle = {Proceedings of the 17th ACM Conference on Computer Supported Cooperative Work \& Social Computing},
pages = {1188–1199},
numpages = {12},
location = {Baltimore, Maryland, USA},
series = {CSCW '14}
}

@article{agapieCrowdsourcingFieldCase2015,
	title = {Crowdsourcing in the {Field}: {A} {Case} {Study} {Using} {Local} {Crowds} for {Event} {Reporting}},
	volume = {3},
	copyright = {Copyright (c) 2015 Proceedings of the AAAI Conference on Human Computation and Crowdsourcing},
	issn = {2769-1349},
	shorttitle = {Crowdsourcing in the {Field}},
	url = {https://ojs.aaai.org/index.php/HCOMP/article/view/13235},
	doi = {10.1609/hcomp.v3i1.13235},
	language = {en},
	urldate = {2025-12-04},
	journal = {Proceedings of the AAAI Conference on Human Computation and Crowdsourcing},
	author = {Agapie, Elena and Teevan, Jaime and Monroy-Hernández, Andrés},
	month = sep,
	year = {2015},
	pages = {2--11},
}

@article{godelModeratingMobEvaluating2021,
	title = {Moderating with the {Mob}: {Evaluating} the {Efficacy} of {Real}-{Time} {Crowdsourced} {Fact}-{Checking}},
	volume = {1},
	copyright = {Copyright (c) 2021},
	issn = {2770-3142},
	shorttitle = {Moderating with the {Mob}},
	url = {https://www.tsjournal.org/index.php/jots/article/view/15},
	doi = {10.54501/jots.v1i1.15},
	language = {en},
	number = {1},
	urldate = {2025-12-04},
	journal = {Journal of Online Trust and Safety},
	author = {Godel, William and Sanderson, Zeve and Aslett, Kevin and Nagler, Jonathan and Bonneau, Richard and Persily, Nathaniel and Tucker, Joshua A.},
	month = oct,
	year = {2021},
}

@article{martelMisinformationWarningLabels2023,
	title = {Misinformation warning labels are widely effective: {A} review of warning effects and their moderating features},
	volume = {54},
	issn = {2352-250X},
	shorttitle = {Misinformation warning labels are widely effective},
	url = {https://www.sciencedirect.com/science/article/pii/S2352250X23001550},
	doi = {10.1016/j.copsyc.2023.101710},
	urldate = {2025-09-12},
	journal = {Current Opinion in Psychology},
	author = {Martel, Cameron and Rand, David G.},
	month = dec,
	year = {2023},
}

@inproceedings{gamageLabelingSyntheticContent2025,
author = {Gamage, Dilrukshi and Sewwandi, Dilki and Zhang, Min and Bandara, Arosha K},
title = {Labeling Synthetic Content: User Perceptions of Label Designs for AI-Generated Content on Social Media},
year = {2025},
isbn = {9798400713941},
publisher = {Association for Computing Machinery},
address = {New York, NY, USA},
url = {https://doi.org/10.1145/3706598.3713171},
doi = {10.1145/3706598.3713171},
booktitle = {Proceedings of the 2025 CHI Conference on Human Factors in Computing Systems},
articleno = {814},
numpages = {29},
location = {Yokohama, Japan},
series = {CHI '25}
}

@inproceedings{epsteinWillCrowdGame2020,
author = {Epstein, Ziv and Pennycook, Gordon and Rand, David},
title = {Will the Crowd Game the Algorithm? Using Layperson Judgments to Combat Misinformation on Social Media by Downranking Distrusted Sources},
year = {2020},
isbn = {9781450367080},
publisher = {Association for Computing Machinery},
address = {New York, NY, USA},
url = {https://doi.org/10.1145/3313831.3376232},
doi = {10.1145/3313831.3376232},
booktitle = {Proceedings of the 2020 CHI Conference on Human Factors in Computing Systems},
pages = {1–11},
numpages = {11},
location = {Honolulu, HI, USA},
series = {CHI '20}
}

@techreport{johnsonFactsAllegationsPolitical2023,
	title = {The {Facts} {Behind} {Allegations} of {Political} {Bias} on {Social} {Media}},
	url = {https://itif.org/publications/2023/10/26/the-facts-behind-allegations-of-political-bias-on-social-media/},
	language = {en},
	urldate = {2025-09-11},
	author = {Johnson, Ash},
	month = oct,
	year = {2023},
}

@inproceedings{fanDigitalJuriesCivicsOriented2020,
author = {Fan, Jenny and Zhang, Amy X.},
title = {Digital Juries: A Civics-Oriented Approach to Platform Governance},
year = {2020},
isbn = {9781450367080},
publisher = {Association for Computing Machinery},
address = {New York, NY, USA},
url = {https://doi.org/10.1145/3313831.3376293},
doi = {10.1145/3313831.3376293},
booktitle = {Proceedings of the 2020 CHI Conference on Human Factors in Computing Systems},
pages = {1–14},
numpages = {14},
location = {Honolulu, HI, USA},
series = {CHI '20}
}

@article{elliottElonMusksMain,
	title = {Elon {Musk}’s {Main} {Tool} for {Fighting} {Disinformation} on {X} {Is} {Making} the {Problem} {Worse}, {Insiders} {Claim}},
	issn = {1059-1028},
	url = {https://www.wired.com/story/x-community-notes-disinformation/},
	language = {en-US},
	urldate = {2025-09-11},
	journal = {Wired},
	author = {Elliott, Vittoria},
	month = oct,
	year = {2023},
}

@article{cotterFactCheckingCrisisCOVID192022,
	title = {Fact-{Checking} the {Crisis}: {COVID}-19, {Infodemics}, and the {Platformization} of {Truth}},
	volume = {8},
	issn = {2056-3051},
	shorttitle = {Fact-{Checking} the {Crisis}},
	url = {https://doi.org/10.1177/20563051211069048},
	doi = {10.1177/20563051211069048},
	language = {EN},
	number = {1},
	urldate = {2025-09-10},
	journal = {Social Media + Society},
	author = {Cotter, Kelley and DeCook, Julia R. and Kanthawala, Shaheen},
	month = jan,
	year = {2022},
}

@article{heSurveyRoleCrowds2025,
	title = {A {Survey} on the {Role} of {Crowds} in {Combating} {Online} {Misinformation}: {Annotators}, {Evaluators}, and {Creators}},
	volume = {19},
	issn = {1556-4681, 1556-472X},
	shorttitle = {A {Survey} on the {Role} of {Crowds} in {Combating} {Online} {Misinformation}},
	url = {https://dl.acm.org/doi/10.1145/3694980},
	doi = {10.1145/3694980},
	language = {en},
	number = {1},
	urldate = {2025-09-10},
	journal = {ACM Transactions on Knowledge Discovery from Data},
	author = {He, Bing and Hu, Yibo and Lee, Yeon-Chang and Oh, Soyoung and Verma, Gaurav and Kumar, Srijan},
	month = jan,
	year = {2025},
	pages = {1--30},
}

@inproceedings{lampeSlashdotBurnDistributed2004,
author = {Lampe, Cliff and Resnick, Paul},
title = {Slash(dot) and burn: distributed moderation in a large online conversation space},
year = {2004},
isbn = {1581137028},
publisher = {Association for Computing Machinery},
address = {New York, NY, USA},
url = {https://doi.org/10.1145/985692.985761},
doi = {10.1145/985692.985761},
booktitle = {Proceedings of the SIGCHI Conference on Human Factors in Computing Systems},
pages = {543–550},
numpages = {8},
location = {Vienna, Austria},
series = {CHI '04}
}

@article{seeringReconsideringSelfModerationRole2020,
	title = {Reconsidering {Self}-{Moderation}: the {Role} of {Research} in {Supporting} {Community}-{Based} {Models} for {Online} {Content} {Moderation}},
	volume = {4},
	shorttitle = {Reconsidering {Self}-{Moderation}},
	url = {https://dl.acm.org/doi/10.1145/3415178},
	doi = {10.1145/3415178},
	number = {CSCW2},
	urldate = {2025-09-10},
	journal = {Proc. ACM Hum.-Comput. Interact.},
	author = {Seering, Joseph},
	month = oct,
	year = {2020},
	pages = {107:1--107:28},
}

@inproceedings{zhaoEffectsNonEffectsSocial2025,
	address = {New York, NY, USA},
	series = {{CHI} '25},
	title = {The {Effects} and {Non}-{Effects} of {Social} {Sanctions} from {User} {Jury}-{Based} {Content} {Moderation} {Decisions} on {Weibo}},
	isbn = {979-8-4007-1394-1},
	url = {https://dl.acm.org/doi/10.1145/3706598.3713154},
	doi = {10.1145/3706598.3713154},
	urldate = {2025-09-10},
	booktitle = {Proceedings of the 2025 {CHI} {Conference} on {Human} {Factors} in {Computing} {Systems}},
	publisher = {Association for Computing Machinery},
	author = {Zhao, Andy and Hobbs, Will},
	month = apr,
	year = {2025},
	pages = {1--17},
    location = {Yokohama, Japan},
}

@article{zhaoInsightsComparativeStudy2023,
	title = {Insights from a {Comparative} {Study} on the {Variety}, {Velocity}, {Veracity}, and {Viability} of {Crowdsourced} and {Professional} {Fact}-{Checking} {Services}},
	volume = {2},
	copyright = {Copyright (c) 2023 Journal of Online Trust and Safety},
	issn = {2770-3142},
	url = {https://www.tsjournal.org/index.php/jots/article/view/118},
	doi = {10.54501/jots.v2i1.118},
	language = {en},
	number = {1},
	urldate = {2025-09-10},
	journal = {Journal of Online Trust and Safety},
	author = {Zhao, Andy and Naaman, Mor},
	month = sep,
	year = {2023},
}

@article{martelCrowdsCanEffectively2024a,
	title = {Crowds {Can} {Effectively} {Identify} {Misinformation} at {Scale}},
	volume = {19},
	issn = {1745-6916},
	url = {https://doi.org/10.1177/17456916231190388},
	doi = {10.1177/17456916231190388},
	language = {EN},
	number = {2},
	urldate = {2025-09-10},
	journal = {Perspectives on Psychological Science},
	author = {Martel, Cameron and Allen, Jennifer and Pennycook, Gordon and Rand, David G.},
	month = mar,
	year = {2024},
	pages = {477--488},
}

@inproceedings{mosleh21,
author = {Mosleh, Mohsen and Martel, Cameron and Eckles, Dean and Rand, David},
title = {Perverse Downstream Consequences of Debunking: Being Corrected by Another User for Posting False Political News Increases Subsequent Sharing of Low Quality, Partisan, and Toxic Content in a Twitter Field Experiment},
year = {2021},
isbn = {9781450380966},
publisher = {Association for Computing Machinery},
address = {New York, NY, USA},
url = {https://doi.org/10.1145/3411764.3445642},
doi = {10.1145/3411764.3445642},
booktitle = {Proceedings of the 2021 CHI Conference on Human Factors in Computing Systems},
articleno = {182},
numpages = {13},
location = {Yokohama, Japan},
series = {CHI '21}
}

@article{allenScalingFactcheckingUsing2021,
	title = {Scaling up fact-checking using the wisdom of crowds},
	volume = {7},
	url = {https://www.science.org/doi/full/10.1126/sciadv.abf4393},
	doi = {10.1126/sciadv.abf4393},
	number = {36},
	urldate = {2025-09-10},
	journal = {Science Advances},
	author = {Allen, Jennifer and Arechar, Antonio A. and Pennycook, Gordon and Rand, David G.},
	month = sep,
	year = {2021},
}

@article{renault25,
author = {Thomas Renault  and Mohsen Mosleh  and David G. Rand },
title = {Republicans are flagged more often than Democrats for sharing misinformation on X’s Community Notes},
journal = {Proceedings of the National Academy of Sciences},
volume = {122},
number = {25},
year = {2025},
doi = {10.1073/pnas.2502053122},
URL = {https://www.pnas.org/doi/abs/10.1073/pnas.2502053122},
}

@misc{mohammadi2025birdwatchcommunitynotestwitter,
      title={From Birdwatch to Community Notes, from Twitter to X: four years of community-based content moderation}, 
      author={Saeedeh Mohammadi and Narges Chinichian and Hannah Doyal and Kristina Skutilova and Hao Cui and Michele d'Errico and Siobhan Grayson and Taha Yasseri},
      year={2025},
      eprint={2510.09585},
      archivePrefix={arXiv},
      url={https://arxiv.org/abs/2510.09585}, 
}

@misc{martelPoliticalMotivesHelp2025,
	title = {Political motives help rather than hinder crowdsourced fact-checking},
	url = {https://osf.io/preprints/psyarxiv/8fhxz_v1/},
	doi = {10.31234/osf.io/8fhxz_v1},
	urldate = {2025-12-03},
	publisher = {PsyArXiv},
	author = {Martel, Cameron and Allen, Jennifer N L and Pennycook, Gordon and Rand, David G},
	month = jun,
	year = {2025},
}

@article{fraxanetUnpackingPolarizationAntagonism2024c,
	title = {Unpacking polarization: {Antagonism} and alignment in signed networks of online interaction},
	volume = {3},
	issn = {2752-6542},
	shorttitle = {Unpacking polarization},
	url = {https://doi.org/10.1093/pnasnexus/pgae276},
	doi = {10.1093/pnasnexus/pgae276},
	number = {12},
	urldate = {2025-12-03},
	journal = {PNAS Nexus},
	author = {Fraxanet, Emma and Pellert, Max and Schweighofer, Simon and Gómez, Vicenç and Garcia, David},
	month = dec,
	year = {2024},
	pages = {pgae276},
}

@article{pennycookFightingMisinformationSocial2019,
	title = {Fighting misinformation on social media using crowdsourced judgments of news source quality},
	volume = {116},
	url = {https://www.pnas.org/doi/abs/10.1073/pnas.1806781116},
	doi = {10.1073/pnas.1806781116},
	number = {7},
	urldate = {2025-09-10},
	journal = {Proceedings of the National Academy of Sciences},
	author = {Pennycook, Gordon and Rand, David G.},
	month = feb,
	year = {2019},
	pages = {2521--2526},
}

@article{micallefTrueFalseStudying2022,
author = {Micallef, Nicholas and Armacost, Vivienne and Memon, Nasir and Patil, Sameer},
title = {True or False: Studying the Work Practices of Professional Fact-Checkers},
year = {2022},
issue_date = {April 2022},
publisher = {Association for Computing Machinery},
address = {New York, NY, USA},
volume = {6},
number = {CSCW1},
url = {https://doi.org/10.1145/3512974},
doi = {10.1145/3512974},
journal = {Proc. ACM Hum.-Comput. Interact.},
month = apr,
articleno = {127},
numpages = {44}
}

@inproceedings{tanakaWhoDoesNot2023,
author = {Tanaka, Yuko and Inuzuka, Miwa and Arai, Hiromi and Takahashi, Yoichi and Kukita, Minao and Inui, Kentaro},
title = {Who Does Not Benefit from Fact-checking Websites? A Psychological Characteristic Predicts the Selective Avoidance of Clicking Uncongenial Facts},
year = {2023},
isbn = {9781450394215},
publisher = {Association for Computing Machinery},
address = {New York, NY, USA},
url = {https://doi.org/10.1145/3544548.3580826},
doi = {10.1145/3544548.3580826},
booktitle = {Proceedings of the 2023 CHI Conference on Human Factors in Computing Systems},
articleno = {664},
numpages = {17},
location = {Hamburg, Germany},
series = {CHI '23}
}

@article{claytonRealSolutionsFake2020,
	title = {Real {Solutions} for {Fake} {News}? {Measuring} the {Effectiveness} of {General} {Warnings} and {Fact}-{Check} {Tags} in {Reducing} {Belief} in {False} {Stories} on {Social} {Media}},
	volume = {42},
	issn = {1573-6687},
	shorttitle = {Real {Solutions} for {Fake} {News}?},
	url = {https://doi.org/10.1007/s11109-019-09533-0},
	doi = {10.1007/s11109-019-09533-0},
	language = {en},
	number = {4},
	urldate = {2025-09-10},
	journal = {Political Behavior},
	author = {Clayton, Katherine and Blair, Spencer and Busam, Jonathan A. and Forstner, Samuel and Glance, John and Green, Guy and Kawata, Anna and Kovvuri, Akhila and Martin, Jonathan and Morgan, Evan and Sandhu, Morgan and Sang, Rachel and Scholz-Bright, Rachel and Welch, Austin T. and Wolff, Andrew G. and Zhou, Amanda and Nyhan, Brendan},
	month = dec,
	year = {2020},
	pages = {1073--1095},
}

@article{junejaHumanTechnologicalInfrastructures2022,
	title = {Human and {Technological} {Infrastructures} of {Fact}-checking},
	volume = {6},
	url = {https://dl.acm.org/doi/10.1145/3555143},
	doi = {10.1145/3555143},
	number = {CSCW2},
	urldate = {2025-09-10},
	journal = {Proc. ACM Hum.-Comput. Interact.},
	author = {Juneja, Prerna and Mitra, Tanushree},
	month = nov,
	year = {2022},
	pages = {418:1--418:36},
}

@inproceedings{warrenShowMeWork2025,
author = {Warren, Greta and Shklovski, Irina and Augenstein, Isabelle},
title = {Show Me the Work: Fact-Checkers' Requirements for Explainable Automated Fact-Checking},
year = {2025},
isbn = {9798400713941},
publisher = {Association for Computing Machinery},
address = {New York, NY, USA},
url = {https://doi.org/10.1145/3706598.3713277},
booktitle = {Proceedings of the 2025 CHI Conference on Human Factors in Computing Systems},
articleno = {421},
numpages = {21},
location = {Yokohama, Japan},
}

@article{hassanQuestAutomateFactChecking,
	title = {The {Quest} to {Automate} {Fact}-{Checking}},
	language = {en},
	author = {Hassan, Naeemul and Adair, Bill and Hamilton, James T and Li, Chengkai and Tremayne, Mark and Yang, Jun and Yu, Cong},
    year = {2015},
    journal = {Proceedings of the 2015 Computation + Journalism Symposium}
}

@article{guoSurveyAutomatedFactChecking2022,
	title = {A {Survey} on {Automated} {Fact}-{Checking}},
	volume = {10},
	issn = {2307-387X},
	url = {https://doi.org/10.1162/tacl_a_00454},
	doi = {10.1162/tacl_a_00454},
	urldate = {2025-09-10},
	journal = {Transactions of the Association for Computational Linguistics},
	author = {Guo, Zhijiang and Schlichtkrull, Michael and Vlachos, Andreas},
	month = feb,
	year = {2022},
	pages = {178--206},
}

@inproceedings{hassanDetectingCheckworthyFactual2015,
author = {Hassan, Naeemul and Li, Chengkai and Tremayne, Mark},
title = {Detecting Check-worthy Factual Claims in Presidential Debates},
year = {2015},
isbn = {9781450337946},
publisher = {Association for Computing Machinery},
address = {New York, NY, USA},
url = {https://doi.org/10.1145/2806416.2806652},
doi = {10.1145/2806416.2806652},
booktitle = {Proceedings of the 24th ACM International on Conference on Information and Knowledge Management},
pages = {1835–1838},
numpages = {4},
location = {Melbourne, Australia},
series = {CIKM '15}
}

@article{Kankham18062025,
author = {Sarawut Kankham and Jian-Ren Hou},
title = {Community Notes vs. Related Articles: Assessing Real-World Integrated Counter-Rumor Features in Response to Different Rumor Types on Social Media},
journal = {International Journal of Human–Computer Interaction},
volume = {41},
number = {12},
pages = {7711--7725},
year = {2025},
publisher = {Taylor \& Francis},
doi = {10.1080/10447318.2024.2400389},
URL = { https://doi.org/10.1080/10447318.2024.2400389}
}

@article{youngFactCheckingEffectivenessFunction2018,
	title = {Fact-{Checking} {Effectiveness} as a {Function} of {Format} and {Tone}: {Evaluating} {FactCheck}.org and {FlackCheck}.org},
	volume = {95},
	issn = {1077-6990},
	shorttitle = {Fact-{Checking} {Effectiveness} as a {Function} of {Format} and {Tone}},
	url = {https://doi.org/10.1177/1077699017710453},
	doi = {10.1177/1077699017710453},
	language = {EN},
	number = {1},
	urldate = {2025-09-10},
	journal = {Journalism \& Mass Communication Quarterly},
	author = {Young, Dannagal G. and Jamieson, Kathleen Hall and Poulsen, Shannon and Goldring, Abigail},
	month = mar,
	year = {2018},
	pages = {49--75},
}

@article{walterFactCheckingMetaAnalysisWhat2020,
	title = {Fact-{Checking}: {A} {Meta}-{Analysis} of {What} {Works} and for {Whom}},
	volume = {37},
	issn = {1058-4609},
	shorttitle = {Fact-{Checking}},
	url = {https://doi.org/10.1080/10584609.2019.1668894},
	doi = {10.1080/10584609.2019.1668894},
	number = {3},
	urldate = {2025-09-10},
	journal = {Political Communication},
	author = {Walter, Nathan and Cohen, Jonathan and Holbert, R. Lance and Morag, Yasmin},
	month = may,
	year = {2020},
	pages = {350--375},
}

@misc{wojcikBirdwatchCrowdWisdom2022b,
	title = {Birdwatch: {Crowd} {Wisdom} and {Bridging} {Algorithms} can {Inform} {Understanding} and {Reduce} the {Spread} of {Misinformation}},
	shorttitle = {Birdwatch},
	url = {http://arxiv.org/abs/2210.15723},
	doi = {10.48550/arXiv.2210.15723},
	urldate = {2025-09-04},
	publisher = {arXiv},
	author = {Wojcik, Stefan and Hilgard, Sophie and Judd, Nick and Mocanu, Delia and Ragain, Stephen and Hunzaker, M. B. Fallin and Coleman, Keith and Baxter, Jay},
	month = oct,
	year = {2022},
}

@misc{pilarskiCommunityNotesVs2023,
	title = {Community {Notes} vs. {Snoping}: {How} the {Crowd} {Selects} {Fact}-{Checking} {Targets} on {Social} {Media}},
	shorttitle = {Community {Notes} vs. {Snoping}},
	url = {http://arxiv.org/abs/2305.09519},
	doi = {10.48550/arXiv.2305.09519},
	urldate = {2025-08-07},
	publisher = {arXiv},
	author = {Pilarski, Moritz and Solovev, Kirill and Pröllochs, Nicolas},
	month = sep,
	year = {2023},
}

@misc{chuai2025requestnoterequestfunction,
      title={Request a Note: How the Request Function Shapes X's Community Notes System}, 
      author={Yuwei Chuai and Shuning Zhang and Ziming Wang and Xin Yi and Mohsen Mosleh and Gabriele Lenzini},
      year={2025},
      eprint={2509.09956},
      archivePrefix={arXiv},
      url={https://arxiv.org/abs/2509.09956}, 
}

@article{diffusion21,
author = {Drolsbach, Chiara Patricia and Pr\"{o}llochs, Nicolas},
title = {Diffusion of Community Fact-Checked Misinformation on Twitter},
year = {2023},
issue_date = {October 2023},
publisher = {Association for Computing Machinery},
address = {New York, NY, USA},
volume = {7},
number = {CSCW2},
url = {https://doi.org/10.1145/3610058},
doi = {10.1145/3610058},
journal = {Proc. ACM Hum.-Comput. Interact.},
month = oct,
articleno = {267},
numpages = {22}
}

@article{slaughter25,
author = {Isaac Slaughter  and Axel Peytavin  and Johan Ugander  and Martin Saveski },
title = {Community notes reduce engagement with and diffusion of false information online},
journal = {Proceedings of the National Academy of Sciences},
volume = {122},
number = {38},
pages = {e2503413122},
year = {2025},
doi = {10.1073/pnas.2503413122},
URL = {https://www.pnas.org/doi/abs/10.1073/pnas.2503413122},
}

@article{rollout21,
author = {Chuai, Yuwei and Tian, Haoye and Pr\"{o}llochs, Nicolas and Lenzini, Gabriele},
title = {Did the Roll-Out of Community Notes Reduce Engagement With Misinformation on X/Twitter?},
year = {2024},
issue_date = {November 2024},
publisher = {Association for Computing Machinery},
address = {New York, NY, USA},
volume = {8},
number = {CSCW2},
url = {https://doi.org/10.1145/3686967},
doi = {10.1145/3686967},
journal = {Proc. ACM Hum.-Comput. Interact.},
month = nov,
articleno = {428},
numpages = {52}
}

@misc{pröllochs2021communitybasedfactcheckingtwittersbirdwatch,
      title={Community-Based Fact-Checking on Twitter's Birdwatch Platform}, 
      author={Nicolas Pröllochs},
      year={2021},
      eprint={2104.07175},
      archivePrefix={arXiv},
      url={https://arxiv.org/abs/2104.07175}, 
}

@misc{chuaiCommunitybasedFactcheckingReduces2024a,
	title = {Community-based fact-checking reduces the spread of misleading posts on social media},
	url = {http://arxiv.org/abs/2409.08781},
	doi = {10.48550/arXiv.2409.08781},
	urldate = {2025-08-07},
	publisher = {arXiv},
	author = {Chuai, Yuwei and Pilarski, Moritz and Renault, Thomas and Restrepo-Amariles, David and Troussel-Clément, Aurore and Lenzini, Gabriele and Pröllochs, Nicolas},
	month = sep,
	year = {2024},
}

@article{nixMetaEndsFactchecking2025,
	title = {Meta ends fact-checking, drawing praise from {Trump}},
	issn = {0190-8286},
	url = {https://www.washingtonpost.com/technology/2025/01/07/meta-factchecking-zuckerberg/},
	language = {en-US},
	urldate = {2025-09-07},
	journal = {The Washington Post},
	author = {Nix, Naomi and Oremus, Will and Gregg, Aaron},
	month = jan,
	year = {2025},
}

@misc{TestingNewWays,
	title = {Testing new ways to offer viewers more context and information on videos},
	url = {https://blog.youtube/news-and-events/new-ways-to-offer-viewers-more-context/},
	language = {en-us},
	urldate = {2025-09-07},
	journal = {blog.youtube},
        author = "The Youtube Team",
        month = jun,
        year = 2024
}

@article{grimmelmann2015virtues,
  title={The virtues of moderation},
  author={Grimmelmann, James},
  journal={Yale JL \& Tech.},
  volume={17},
  pages={42},
  year={2015},
  publisher={HeinOnline}
}

@misc{TestingNewFeature,
	title = {Testing a new feature to enhance content on {TikTok}},
	url = {https://newsroom.tiktok.com/footnotes},
	language = {en},
	urldate = {2025-09-07},
	journal = {Newsroom {\textbar} TikTok},
	author = {Presser, Adam},
	month = apr,
	year = {2025},
}

@misc{heatheraMoreSpeechFewer2025,
	title = {More {Speech} and {Fewer} {Mistakes}},
	url = {https://about.fb.com/news/2025/01/meta-more-speech-fewer-mistakes/},
	language = {en-US},
	urldate = {2025-09-07},
	journal = {Meta Newsroom},
	author = {Joel Kaplan},
	month = jan,
	year = {2025},
}

@inproceedings{cikm22,
author = {Saeed, Mohammed and Traub, Nicolas and Nicolas, Maelle and Demartini, Gianluca and Papotti, Paolo},
title = {Crowdsourced Fact-Checking at Twitter: How Does the Crowd Compare With Experts?},
year = {2022},
isbn = {9781450392365},
publisher = {Association for Computing Machinery},
address = {New York, NY, USA},
url = {https://doi.org/10.1145/3511808.3557279},
doi = {10.1145/3511808.3557279},
booktitle = {Proceedings of the 31st ACM International Conference on Information \& Knowledge Management},
pages = {1736–1746},
numpages = {11},
location = {Atlanta, GA, USA},
series = {CIKM '22}
}

@article{Hameleers18052024,
	title = {Why do social media users accept, doubt or resist corrective information? {A} qualitative analysis of comments in response to corrective information on social media},
	volume = {25},
	url = {https://doi.org/10.1080/1461670X.2024.2340591},
	doi = {10.1080/1461670X.2024.2340591},
	number = {7},
	journal = {Journalism Studies},
	author = {Hameleers, Michael},
	year = {2024},
	pages = {776--793},
}

@misc{renaultCollaborativelyAddingContext2024a,
	title = {Collaboratively adding context to social media posts reduces the sharing of false news},
	url = {http://arxiv.org/abs/2404.02803},
	doi = {10.48550/arXiv.2404.02803},
	urldate = {2025-08-07},
	publisher = {arXiv},
	author = {Renault, Thomas and Amariles, David Restrepo and Troussel, Aurore},
	month = apr,
	year = {2024},
}

@misc{IntroducingBirdwatchCommunitybaseda,
	title = {Introducing {Birdwatch}, a community-based approach to misinformation},
	url = {https://blog.x.com/en_us/topics/product/2021/introducing-birdwatch-a-community-based-approach-to-misinformation},
	language = {en-us},
	urldate = {2025-09-04},
        author="Keith Coleman",
        year="2021",
        month="January"
}

@misc{BeginningTodayCommunity2022,
	type = {Tweet},
	title = {Beginning today, {Community} {Notes} are visible around the world},
	url = {https://x.com/CommunityNotes/status/1601753552476438528},
	language = {en},
	urldate = {2025-09-08},
	journal = {Twitter},
	author = {{@CommunityNotes}},
	month = dec,
	year = {2022},
}

@misc{BuildingBetterBirdwatch,
	title = {Building a better {Birdwatch}},
	url = {https://blog.x.com/en_us/topics/company/2022/building-a-better-birdwatch},
	language = {en-us},
	urldate = {2025-09-08},
        author="Keith Coleman",
        year="2022",
        month=mar
}

@inproceedings{hawkeye,
author = {Mujumdar, Rohit and Kumar, Srijan},
title = {HawkEye: a robust reputation system for community-based counter-misinformation},
year = {2022},
isbn = {9781450391283},
publisher = {Association for Computing Machinery},
address = {New York, NY, USA},
url = {https://doi.org/10.1145/3487351.3488343},
doi = {10.1145/3487351.3488343},
booktitle = {Proceedings of the 2021 IEEE/ACM International Conference on Advances in Social Networks Analysis and Mining},
pages = {188–192},
numpages = {5},
location = {Virtual Event, Netherlands},
series = {ASONAM '21}
}

@inproceedings{franzmeyer2024hellofreshllmevaluationsstreams,
    title = "{H}ello{F}resh: {LLM} Evalutions on Streams of Real-World Human Editorial Actions across {X} Community Notes and {W}ikipedia edits",
    author = "Franzmeyer, Tim  and
      Shtedritski, Aleksandar  and
      Albanie, Samuel  and
      Torr, Philip  and
      Henriques, Joao F.  and
      Foerster, Jakob",
    editor = "Ku, Lun-Wei  and
      Martins, Andre  and
      Srikumar, Vivek",
    booktitle = "Findings of the Association for Computational Linguistics: ACL 2024",
    month = aug,
    year = "2024",
    address = "Bangkok, Thailand",
    publisher = "Association for Computational Linguistics",
    url = "https://aclanthology.org/2024.findings-acl.754/",
    doi = "10.18653/v1/2024.findings-acl.754",
    pages = "12702--12716",
}

@inproceedings{allen22,
author = {Allen, Jennifer and Martel, Cameron and Rand, David G},
title = {Birds of a feather don’t fact-check each other: Partisanship and the evaluation of news in Twitter’s Birdwatch crowdsourced fact-checking program},
year = {2022},
isbn = {9781450391573},
publisher = {Association for Computing Machinery},
address = {New York, NY, USA},
url = {https://doi.org/10.1145/3491102.3502040},
doi = {10.1145/3491102.3502040},
booktitle = {Proceedings of the 2022 CHI Conference on Human Factors in Computing Systems},
articleno = {245},
numpages = {19},
location = {New Orleans, LA, USA},
series = {CHI '22}
}

@inproceedings{drawsEffectsCrowdWorker2022,
author = {Draws, Tim and La Barbera, David and Soprano, Michael and Roitero, Kevin and Ceolin, Davide and Checco, Alessandro and Mizzaro, Stefano},
title = {The Effects of Crowd Worker Biases in Fact-Checking Tasks},
year = {2022},
isbn = {9781450393522},
publisher = {Association for Computing Machinery},
address = {New York, NY, USA},
url = {https://doi.org/10.1145/3531146.3534629},
doi = {10.1145/3531146.3534629},
booktitle = {Proceedings of the 2022 ACM Conference on Fairness, Accountability, and Transparency},
pages = {2114–2124},
numpages = {11},
location = {Seoul, Republic of Korea},
series = {FAccT '22}
}

@preprint{arjmandilari2025threatssustainabilitycommunitynotes,
      title={Threats to the sustainability of Community Notes on X}, 
      author={Zahra Arjmandi-Lari and Alexios Mantzarlis and Tom Stafford},
      year={2025},
      eprint={2510.00650},
      archivePrefix={arXiv},
      url={https://arxiv.org/abs/2510.00650}, 
}

@misc{solovevReferencesUnbiasedSources2025a,
	title = {References to unbiased sources increase the helpfulness of community fact-checks},
	url = {http://arxiv.org/abs/2503.10560},
	doi = {10.48550/arXiv.2503.10560},
	urldate = {2025-08-07},
	publisher = {arXiv},
	author = {Solovev, Kirill and Pröllochs, Nicolas},
	month = mar,
	year = {2025},
}

@incollection{macleanQuestionsOptionsCriteria1996,
	title = {Questions, {Options}, and {Criteria}: {Elements} of {Design} {Space} {Analysis}},
	shorttitle = {Questions, {Options}, and {Criteria}},
	booktitle = {Design {Rationale}},
	publisher = {CRC Press},
	author = {MacLean, Allan and Young, Richard M. and Bellotti, Victoria M. E. and Moran, Thomas P.},
	year = {1996},
}

@inproceedings{shusas24,
author = {Shusas, Erica and Forte, Andrea},
title = {Trust and Transparency: An Exploratory Study on Emerging Adults' Interpretations of Credibility Indicators on Social Media Platforms},
year = {2024},
isbn = {9798400703317},
publisher = {Association for Computing Machinery},
address = {New York, NY, USA},
url = {https://doi.org/10.1145/3613905.3650801},
doi = {10.1145/3613905.3650801},
booktitle = {Extended Abstracts of the CHI Conference on Human Factors in Computing Systems},
articleno = {343},
numpages = {7},
location = {Honolulu, HI, USA},
series = {CHI EA '24}
}

@book{krautBuildingSuccessfulOnline2012,
	title = {Building {Successful} {Online} {Communities}: {Evidence}-{Based} {Social} {Design}},
	isbn = {978-0-262-29739-4},
	shorttitle = {Building {Successful} {Online} {Communities}},
	language = {en},
	publisher = {MIT Press},
	author = {Kraut, Robert E. and Resnick, Paul},
	month = mar,
	year = {2012},
}

@inproceedings{borenstein2025,
    title = "Can Community Notes Replace Professional Fact-Checkers?",
    author = "Borenstein, Nadav  and
      Warren, Greta  and
      Elliott, Desmond  and
      Augenstein, Isabelle",
    editor = "Che, Wanxiang  and
      Nabende, Joyce  and
      Shutova, Ekaterina  and
      Pilehvar, Mohammad Taher",
    booktitle = "Proceedings of the 63rd Annual Meeting of the Association for Computational Linguistics (Volume 2: Short Papers)",
    month = jul,
    year = "2025",
    address = "Vienna, Austria",
    publisher = "Association for Computational Linguistics",
    url = "https://aclanthology.org/2025.acl-short.42/",
    doi = "10.18653/v1/2025.acl-short.42",
    pages = "535--552",
    ISBN = "979-8-89176-252-7",
}

@misc{bobek2025communityfactchecksbreakfollower,
      title={Community Fact-Checks Do Not Break Follower Loyalty}, 
      author={Michelle Bobek and Nicolas Pröllochs},
      year={2025},
      eprint={2505.10254},
      archivePrefix={arXiv},
      url={https://arxiv.org/abs/2505.10254}, 
}

@inproceedings{cham,
	address = {Cham},
	title = {Who watches the birdwatchers? {Sociotechnical} vulnerabilities in twitter's content contextualisation},
	isbn = {978-3-031-10183-0},
	booktitle = {Socio-technical aspects in security},
	publisher = {Springer International Publishing},
	author = {Benjamin, Garfield},
	editor = {Parkin, Simon and Viganò, Luca},
	year = {2022},
	pages = {3--23},
}

@misc{dekeulenaarTwitterDemotionCommunity2025,
	address = {Rochester, NY},
	type = {{SSRN} {Scholarly} {Paper}},
	title = {From {Twitter} to {X}: demotion, community notes and the apparent shift from adjudication to consensus-building},
	shorttitle = {From {Twitter} to {X}},
	url = {https://papers.ssrn.com/abstract=5165083},
	doi = {10.2139/ssrn.5165083},
	language = {en},
	urldate = {2025-12-03},
	publisher = {Social Science Research Network},
	author = {de Keulenaar, Emillie},
	month = mar,
	year = {2025},
}

@misc{colemanLimitingFactorsEffectiveness2023,
	title = {Limiting {Factors} in the {Effectiveness} of {Crowd}-{Sourced} {Labeling} for {Combating} {Misinformation}},
	url = {https://osf.io/preprints/socarxiv/ahm27_v1/},
	doi = {10.31235/osf.io/ahm27},
	urldate = {2025-12-03},
	publisher = {SocArXiv},
	author = {Bak-Coleman, Joseph B},
	month = apr,
	year = {2023},
}

@article{wangEfficiencyCommunityBasedContent2024b,
	title = {Efficiency of {Community}-{Based} {Content} {Moderation} {Mechanisms}: {A} {Discussion} {Focused} on {Birdwatch}},
	volume = {33},
	issn = {1572-9907},
	url = {https://doi.org/10.1007/s10726-024-09881-1},
	doi = {10.1007/s10726-024-09881-1},
	number = {3},
	journal = {Group Decision and Negotiation},
	author = {Wang, Chenlong and Lucas, Pablo},
	month = jun,
	year = {2024},
	pages = {673--709},
}

@article{Wirtschafter_Majumder_2023,
	title = {Future challenges for online, crowdsourced content moderation: {Evidence} from twitter’s community notes},
	volume = {2},
	url = {https://www.tsjournal.org/index.php/jots/article/view/139},
	doi = {10.54501/jots.v2i1.139},
	number = {1},
	journal = {Journal of Online Trust and Safety},
	author = {Wirtschafter, Valerie and Majumder, Sharanya},
	month = sep,
	year = {2023},
}

@misc{borwankarDemocratizationMisinformationMonitoring2022a,
	address = {Rochester, NY},
	type = {{SSRN} {Scholarly} {Paper}},
	title = {Democratization of {Misinformation} {Monitoring}: {The} {Impact} of {Twitter}’s {Birdwatch} {Program}},
	shorttitle = {Democratization of {Misinformation} {Monitoring}},
	url = {https://papers.ssrn.com/abstract=4236756},
	doi = {10.2139/ssrn.4236756},
	language = {en},
	urldate = {2025-12-03},
	publisher = {Social Science Research Network},
	author = {Borwankar, Sameer and Zheng, Jinyang and Kannan, Karthik Natarajan},
	month = oct,
	year = {2022},
}

@article{jama24,
	title = {Characteristics of {X} (formerly twitter) community notes addressing {COVID}-19 vaccine misinformation},
	volume = {331},
	issn = {0098-7484},
	url = {https://doi.org/10.1001/jama.2024.4800},
	doi = {10.1001/jama.2024.4800},
	number = {19},
	journal = {JAMA: The Journal of the American Medical Association},
	author = {Allen, Matthew R. and Desai, Nimit and Namazi, Aiden and Leas, Eric and Dredze, Mark and Smith, Davey M. and Ayers, John W.},
	month = may,
	year = {2024},
	pages = {1670--1672},
}

@article{gao24,
author = {Gao, Yang and Zhang, Maggie Mengqing and Rui, Huaxia},
title = {Can Crowdchecking Curb Misinformation? Evidence from Community Notes},
journal = {Information Systems Research},
year = {2024},
doi = {10.1287/isre.2024.1609},
URL = {https://doi.org/10.1287/isre.2024.1609}
}

@misc{DeepDiveXs,
	title = {A {Deep} {Dive} into {X}’s {Community} {Notes}: {An} {Analysis} of {English} and {Spanish} {Contributions} {Between} 2021 and 2025},
	shorttitle = {A {Deep} {Dive} into {X}’s {Community} {Notes}},
	url = {https://ddia.org/en/a-deep-dive-into-xs-community-notes-report},
	language = {en},
	urldate = {2025-12-03},
	author = {Braga, Roberta and Tardáguila, Cristina and Soares, Marcelo},
	month = jul,
	year = {2025},
}

@misc{kangur2024checkscheckersexploringsource,
      title={Who Checks the Checkers? Exploring Source Credibility in Twitter's Community Notes}, 
      author={Uku Kangur and Roshni Chakraborty and Rajesh Sharma},
      year={2024},
      eprint={2406.12444},
      archivePrefix={arXiv},
      url={https://arxiv.org/abs/2406.12444}, 
}

@inproceedings{kuuse25,
author = {Kuuse, Simon Fox and Kangur, Uku and Chakraborty, Roshni and Sharma, Rajesh},
title = {Crowdsourced Fact-Checking or Biased Commentary? Analyzing Political Bias in Twitter's Community Notes},
year = {2025},
isbn = {9798400713316},
publisher = {Association for Computing Machinery},
address = {New York, NY, USA},
url = {https://doi.org/10.1145/3701716.3717533},
doi = {10.1145/3701716.3717533},
booktitle = {Companion Proceedings of the ACM on Web Conference 2025},
pages = {2661–2669},
numpages = {9},
location = {Sydney NSW, Australia},
series = {WWW '25}
}

@inproceedings{jones22,
author = {Jones, Isaiah and Hecht, Brent and Vincent, Nicholas},
title = {Misleading Tweets and Helpful Notes: Investigating Data Labor by Twitter Birdwatch Users},
year = {2022},
isbn = {9781450391900},
publisher = {Association for Computing Machinery},
address = {New York, NY, USA},
url = {https://doi.org/10.1145/3500868.3559461},
doi = {10.1145/3500868.3559461},
booktitle = {Companion Publication of the 2022 Conference on Computer Supported Cooperative Work and Social Computing},
pages = {68–71},
numpages = {4},
location = {Virtual Event, Taiwan},
series = {CSCW'22 Companion}
}

@misc{xing2025communitynotesdatasetexploringhelpfulness,
      title={COMMUNITYNOTES: A Dataset for Exploring the Helpfulness of Fact-Checking Explanations}, 
      author={Rui Xing and Preslav Nakov and Timothy Baldwin and Jey Han Lau},
      year={2025},
      eprint={2510.24810},
      archivePrefix={arXiv},
      url={https://arxiv.org/abs/2510.24810}, 
}

@inproceedings{
singh2025on,
title={On the Limitations of {LLM}-Synthesized Social Media Misinformation Moderation},
author={Sahajpreet Singh and Jiaying Wu and Svetlana Churina and Kokil Jaidka and Min-Yen Kan},
booktitle={I Can't Believe It's Not Better: Challenges in Applied Deep Learning},
year={2025},
url={https://openreview.net/forum?id=ilz2ghLgzt}
}

@misc{zhang2025commenotessynthesizingorganiccomments,
      title={Commenotes: Synthesizing Organic Comments to Support Community-Based Fact-Checking}, 
      author={Shuning Zhang and Linzhi Wang and Dai Shi and Yuwei Chuai and Jingruo Chen and Yunyi Chen and Yifan Wang and Yating Wang and Xin Yi and Hewu Li},
      year={2025},
      eprint={2509.11052},
      archivePrefix={arXiv},
      url={https://arxiv.org/abs/2509.11052}, 
}

@inproceedings{supernotes,
author = {De, Soham and Bakker, Michiel A. and Baxter, Jay and Saveski, Martin},
title = {Supernotes: Driving Consensus in Crowd-Sourced Fact-Checking},
year = {2025},
isbn = {9798400712746},
publisher = {Association for Computing Machinery},
address = {New York, NY, USA},
url = {https://doi.org/10.1145/3696410.3714934},
doi = {10.1145/3696410.3714934},
booktitle = {Proceedings of the ACM on Web Conference 2025},
pages = {3751–3761},
numpages = {11},
location = {Sydney NSW, Australia},
series = {WWW '25}
}

@article{beyer99,
author = {Beyer, Hugh and Holtzblatt, Karen},
title = {Contextual design},
year = {1999},
issue_date = {Jan./Feb. 1999},
publisher = {Association for Computing Machinery},
address = {New York, NY, USA},
volume = {6},
number = {1},
issn = {1072-5520},
url = {https://doi.org/10.1145/291224.291229},
doi = {10.1145/291224.291229},
journal = {Interactions},
month = jan,
pages = {32–42},
numpages = {11}
}

@article{braunUsingThematicAnalysis2006,
	title = {Using thematic analysis in psychology},
	volume = {3},
	issn = {1478-0887},
	url = {https://doi.org/10.1191/1478088706qp063oa},
	doi = {10.1191/1478088706qp063oa},
	number = {2},
	urldate = {2025-09-10},
	journal = {Qualitative Research in Psychology},
	author = {Braun, Virginia and Clarke, Victoria},
	month = jan,
	year = {2006},
	pages = {77--101},
}

@article{kimDifferentialImpactIndividual2025e,
	title = {Differential impact from individual versus collective misinformation tagging on the diversity of {Twitter} ({X}) information engagement and mobility},
	volume = {16},
	issn = {2041-1723},
	url = {https://doi.org/10.1038/s41467-025-55868-0},
	doi = {10.1038/s41467-025-55868-0},
	number = {1},
	journal = {Nature Communications},
	author = {Kim, Junsol and Wang, Zhao and Shi, Haohan and Ling, Hsin-Keng and Evans, James},
	month = jan,
	year = {2025},
	pages = {973},
}

@inproceedings{kitturFutureCrowdWork2013,
author = {Kittur, Aniket and Nickerson, Jeffrey V. and Bernstein, Michael and Gerber, Elizabeth and Shaw, Aaron and Zimmerman, John and Lease, Matt and Horton, John},
title = {The future of crowd work},
year = {2013},
isbn = {9781450313315},
publisher = {Association for Computing Machinery},
address = {New York, NY, USA},
url = {https://doi.org/10.1145/2441776.2441923},
doi = {10.1145/2441776.2441923},
booktitle = {Proceedings of the 2013 Conference on Computer Supported Cooperative Work},
pages = {1301–1318},
numpages = {18},
location = {San Antonio, Texas, USA},
series = {CSCW '13}
}

@article{bernsteinSoylentWordProcessor2015,
author = {Bernstein, Michael S. and Little, Greg and Miller, Robert C. and Hartmann, Bj\"{o}rn and Ackerman, Mark S. and Karger, David R. and Crowell, David and Panovich, Katrina},
title = {Soylent: a word processor with a crowd inside},
year = {2015},
issue_date = {August 2015},
publisher = {Association for Computing Machinery},
address = {New York, NY, USA},
volume = {58},
number = {8},
issn = {0001-0782},
url = {https://doi.org/10.1145/2791285},
doi = {10.1145/2791285},
journal = {Commun. ACM},
month = jul,
pages = {85–94},
numpages = {10}
}

@misc{wu2025crowdllmaugmentedcommunitynotes,
      title={Beyond the Crowd: LLM-Augmented Community Notes for Governing Health Misinformation}, 
      author={Jiaying Wu and Zihang Fu and Haonan Wang and Fanxiao Li and Min-Yen Kan},
      year={2025},
      eprint={2510.11423},
      archivePrefix={arXiv},
      url={https://arxiv.org/abs/2510.11423}, 
}

@misc{mohammadiAIFeedbackEnhances2025a,
	title = {{AI} {Feedback} {Enhances} {Community}-{Based} {Content} {Moderation} through {Engagement} with {Counterarguments}},
	url = {http://arxiv.org/abs/2507.08110},
	doi = {10.48550/arXiv.2507.08110},
	urldate = {2025-08-07},
	publisher = {arXiv},
	author = {Mohammadi, Saeedeh and Yasseri, Taha},
	month = jul,
	year = {2025},
}

@article{yoon25,
author = {Yoon, Jina and Sathyanarayanan, Shreya and Roesner, Franziska and Zhang, Amy X.},
title = {The Collaborative Practices and Motivations of Online Communities Dedicated to Voluntary Misinformation Response},
year = {2025},
issue_date = {January 2025},
publisher = {Association for Computing Machinery},
address = {New York, NY, USA},
volume = {9},
number = {1},
url = {https://doi.org/10.1145/3701209},
doi = {10.1145/3701209},
journal = {Proc. ACM Hum.-Comput. Interact.},
month = jan,
articleno = {GROUP30},
numpages = {15}
}

@article{drolsbachCommunityNotesIncrease2024a,
	title = {Community notes increase trust in fact-checking on social media},
	volume = {3},
	issn = {2752-6542},
	url = {https://doi.org/10.1093/pnasnexus/pgae217},
	doi = {10.1093/pnasnexus/pgae217},
	number = {7},
	urldate = {2025-09-09},
	journal = {PNAS Nexus},
	author = {Drolsbach, Chiara Patricia and Solovev, Kirill and Pröllochs, Nicolas},
	month = jul,
	year = {2024},
	pages = {pgae217},
}

@article{daileyJournalistsCrowdsourcerersResponding2014,
	title = {Journalists as {Crowdsourcerers}: {Responding} to {Crisis} by {Reporting} with a {Crowd}},
	volume = {23},
	issn = {1573-7551},
	shorttitle = {Journalists as {Crowdsourcerers}},
	url = {https://doi.org/10.1007/s10606-014-9208-z},
	doi = {10.1007/s10606-014-9208-z},
	language = {en},
	number = {4},
	urldate = {2025-12-04},
	journal = {Computer Supported Cooperative Work (CSCW)},
	author = {Dailey, Dharma and Starbird, Kate},
	month = dec,
	year = {2014},
	pages = {445--481},
}

@article{danielQualityControlCrowdsourcing2018,
	title = {Quality {Control} in {Crowdsourcing}: {A} {Survey} of {Quality} {Attributes}, {Assessment} {Techniques}, and {Assurance} {Actions}},
	volume = {51},
	issn = {0360-0300},
	shorttitle = {Quality {Control} in {Crowdsourcing}},
	url = {https://dl.acm.org/doi/10.1145/3148148},
	doi = {10.1145/3148148},
	number = {1},
	urldate = {2025-09-11},
	journal = {ACM Comput. Surv.},
	author = {Daniel, Florian and Kucherbaev, Pavel and Cappiello, Cinzia and Benatallah, Boualem and Allahbakhsh, Mohammad},
	month = jan,
	year = {2018},
	pages = {7:1--7:40},
}

@article{liu23,
	title = {Checking the fact-checkers: {The} role of source type, perceived credibility, and individual differences in fact-checking effectiveness},
	volume = {52},
	url = {https://doi.org/https://doi.org/10.1177/00936502231206419},
	doi = {10.1177/00936502231206419},
	number = {6},
	journal = {Communication Research},
	author = {Liu, Xingyu and Qi, Li and Wang, Laurent and Metzger, Miriam J.},
	year = {2023},
	pages = {719--746},
}

@inproceedings{sharevski2025helps,
  title={" Helps me Take the Post With a Grain of $\{$Salt:$\}$" Soft Moderation Effects on Accuracy Perceptions and Sharing Intentions of Inauthentic Political Content on X},
  author={Sharevski, Filipo and Distler, Verena and Alt, Florian},
  booktitle={34th USENIX Security Symposium (USENIX Security 25)},
  pages={6105--6124},
  year={2025}
}

@inproceedings{jalaliSystematicLiteratureStudies2012,
author = {Jalali, Samireh and Wohlin, Claes},
title = {Systematic literature studies: database searches vs. backward snowballing},
year = {2012},
isbn = {9781450310567},
publisher = {Association for Computing Machinery},
address = {New York, NY, USA},
url = {https://doi.org/10.1145/2372251.2372257},
doi = {10.1145/2372251.2372257},
booktitle = {Proceedings of the ACM-IEEE International Symposium on Empirical Software Engineering and Measurement},
pages = {29–38},
numpages = {10},
location = {Lund, Sweden},
series = {ESEM '12}
}

@inproceedings{chujyo25,
author = {Chujyo, Masaki and Lim, Dongwoo and Tanaka, Mikihito and Toriumi, Fujio},
title = {Community Notes for Preventing the Spread of Misinformation After the Noto Peninsula Earthquake in Japan},
year = {2025},
isbn = {9798400713316},
publisher = {Association for Computing Machinery},
address = {New York, NY, USA},
url = {https://doi.org/10.1145/3701716.3715593},
doi = {10.1145/3701716.3715593},
booktitle = {Companion Proceedings of the ACM on Web Conference 2025},
pages = {938–941},
numpages = {4},
location = {Sydney NSW, Australia},
series = {WWW '25}
}

@misc{razuvayevskaya2025timelinessconsensuscompositioncrowd,
      title={Timeliness, Consensus, and Composition of the Crowd: Community Notes on X}, 
      author={Olesya Razuvayevskaya and Adel Tayebi and Ulrikke Dybdal Sørensen and Kalina Bontcheva and Richard Rogers},
      year={2025},
      eprint={2510.12559},
      archivePrefix={arXiv},
      url={https://arxiv.org/abs/2510.12559}, 
}

@article{matamoros25,
	title = {The importance of centering harm in data infrastructures for ‘soft moderation’: {X}’s {Community} {Notes} as a case study},
	volume = {27},
	url = {https://doi.org/https://doi.org/10.1177/14614448251314399},
	doi = {10.1177/14614448251314399},
	number = {4},
	journal = {New Media \& Society},
	author = {Matamoros-Fernández, Ariadna and Jude, Nadia},
	year = {2025},
	pages = {1986--2011},
}

@misc{liScalingHumanJudgment2025a,
	title = {Scaling {Human} {Judgment} in {Community} {Notes} with {LLMs}},
	url = {http://arxiv.org/abs/2506.24118},
	doi = {10.48550/arXiv.2506.24118},
	urldate = {2025-08-07},
	publisher = {arXiv},
	author = {Li, Haiwen and De, Soham and Revel, Manon and Haupt, Andreas and Miller, Brad and Coleman, Keith and Baxter, Jay and Saveski, Martin and Bakker, Michiel A.},
	month = jun,
	year = {2025},
}

@misc{bouchaudAlgorithmicResolutionCrowdsourced2025b,
	title = {Algorithmic resolution of crowd-sourced moderation on {X} in polarized settings across countries},
	url = {http://arxiv.org/abs/2506.15168},
	doi = {10.48550/arXiv.2506.15168},
	urldate = {2025-08-07},
	publisher = {arXiv},
	author = {Bouchaud, Paul and Ramaciotti, Pedro},
	month = jun,
	year = {2025},
}

@article{pagePRISMA2020Statement2021,
	title = {The {PRISMA} 2020 statement: an updated guideline for reporting systematic reviews},
	volume = {372},
	url = {https://www.bmj.com/content/372/bmj.n71},
	doi = {10.1136/bmj.n71},
	journal = {BMJ (Clinical research ed.)},
	publisher = {BMJ Publishing Group Ltd},
	author = {Page, Matthew J and McKenzie, Joanne E and Bossuyt, Patrick M and Boutron, Isabelle and Hoffmann, Tammy C and Mulrow, Cynthia D and Shamseer, Larissa and Tetzlaff, Jennifer M and Akl, Elie A and Brennan, Sue E and Chou, Roger and Glanville, Julie and Grimshaw, Jeremy M and Hróbjartsson, Asbjørn and Lalu, Manoj M and Li, Tianjing and Loder, Elizabeth W and Mayo-Wilson, Evan and McDonald, Steve and McGuinness, Luke A and Stewart, Lesley A and Thomas, James and Tricco, Andrea C and Welch, Vivian A and Whiting, Penny and Moher, David},
	year = {2021},
}

@inproceedings{wohnHowHandleOnline2017,
author = {Wohn, Donghee Yvette and Fiesler, Casey and Hemphill, Libby and De Choudhury, Munmun and Matias, J. Nathan},
title = {How to Handle Online Risks? Discussing Content Curation and Moderation in Social Media},
year = {2017},
isbn = {9781450346566},
publisher = {Association for Computing Machinery},
address = {New York, NY, USA},
url = {https://doi.org/10.1145/3027063.3051141},
doi = {10.1145/3027063.3051141},
booktitle = {Proceedings of the 2017 CHI Conference Extended Abstracts on Human Factors in Computing Systems},
pages = {1271–1276},
numpages = {6},
location = {Denver, Colorado, USA},
series = {CHI EA '17}
}

@article{vaccaroContestabilityContentModeration2021,
	title = {Contestability {For} {Content} {Moderation}},
	volume = {5},
	url = {https://dl.acm.org/doi/10.1145/3476059},
	doi = {10.1145/3476059},
	number = {CSCW2},
	urldate = {2025-12-04},
	journal = {Proc. ACM Hum.-Comput. Interact.},
	author = {Vaccaro, Kristen and Xiao, Ziang and Hamilton, Kevin and Karahalios, Karrie},
	month = oct,
	year = {2021},
	pages = {318:1--318:28},
}

@article{gillespieNotRecommendReduction2022,
	title = {Do {Not} {Recommend}? {Reduction} as a {Form} of {Content} {Moderation}},
	volume = {8},
	issn = {2056-3051},
	shorttitle = {Do {Not} {Recommend}?},
	url = {https://doi.org/10.1177/20563051221117552},
	doi = {10.1177/20563051221117552},
	language = {EN},
	number = {3},
	urldate = {2025-12-04},
	journal = {Social Media + Society},
	author = {Gillespie, Tarleton},
	month = jul,
	year = {2022},
}

@article{fieslerRedditRulesCharacterizing2018,
	title = {Reddit {Rules}! {Characterizing} an {Ecosystem} of {Governance}},
	volume = {12},
	copyright = {Copyright (c) 2022 Proceedings of the International AAAI Conference on Web and Social Media},
	issn = {2334-0770},
	url = {https://ojs.aaai.org/index.php/ICWSM/article/view/15033},
	doi = {10.1609/icwsm.v12i1.15033},
	language = {en},
	number = {1},
	urldate = {2025-12-04},
	journal = {Proceedings of the International AAAI Conference on Web and Social Media},
	author = {Fiesler, Casey and Jiang, Jialun and McCann, Joshua and Frye, Kyle and Brubaker, Jed},
	month = jun,
	year = {2018},
}

@inproceedings{lloydAIRulesCharacterizing2025,
author = {Lloyd, Travis and Gosciak, Jennah and Nguyen, Tung and Naaman, Mor},
title = {AI Rules? Characterizing Reddit Community Policies Towards AI-Generated Content},
year = {2025},
isbn = {9798400713941},
publisher = {Association for Computing Machinery},
address = {New York, NY, USA},
url = {https://doi.org/10.1145/3706598.3713292},
doi = {10.1145/3706598.3713292},
booktitle = {Proceedings of the 2025 CHI Conference on Human Factors in Computing Systems},
articleno = {9},
numpages = {19},
location = {Yokohama, Japan},
series = {CHI '25}
}

@article{lloyd25,
author = {Lloyd, Travis and Reagle, Joseph and Naaman, Mor},
title = { 'There Has To Be a Lot That We're Missing': Moderating AI-Generated Content on Reddit},
year = {2025},
issue_date = {November 2025},
publisher = {Association for Computing Machinery},
address = {New York, NY, USA},
volume = {9},
number = {7},
url = {https://doi.org/10.1145/3757445},
doi = {10.1145/3757445},
journal = {Proc. ACM Hum.-Comput. Interact.},
month = oct,
articleno = {CSCW264},
numpages = {24}
}

@inproceedings{zhangWhatWeMean2023,
author = {Zhang, Yixuan and Gaggiano, Joseph D and Yongsatianchot, Nutchanon and Suhaimi, Nurul M and Kim, Miso and Sun, Yifan and Griffin, Jacqueline and Parker, Andrea G},
title = {What Do We Mean When We Talk about Trust in Social Media? A Systematic Review},
year = {2023},
isbn = {9781450394215},
publisher = {Association for Computing Machinery},
address = {New York, NY, USA},
url = {https://doi.org/10.1145/3544548.3581019},
doi = {10.1145/3544548.3581019},
booktitle = {Proceedings of the 2023 CHI Conference on Human Factors in Computing Systems},
articleno = {670},
numpages = {22},
location = {Hamburg, Germany},
series = {CHI '23}
}

\appendix

\end{document}